\documentclass[fleqn,nonblindrev,opre]{informs3}
\RequirePackage{tgtermes}
\RequirePackage{newtxtext}
\RequirePackage{newtxmath}
\RequirePackage{bm}
\RequirePackage{endnotes}

\OneAndAHalfSpacedXII 

\usepackage{algorithm}
\usepackage{algpseudocode}
\usepackage{tikz}
\usepackage{framed}
\usepackage{framed}
\usepackage{bm}
\usepackage{tablefootnote}

\def\eps{\epsilon}

\def\X{\mathbf{X}}
\def\Z{\mathbf{Z}}

\def\Q{\mathbf{Q}}
\def\added#1{{#1}}

\def\E{\mathbb{E}}

\def\R{\mathbb{R}}
\def\B{\mathbb{B}}

\def\calA{\mathcal{A}}
\def\calS{\mathcal{S}}
\def\calC{\mathcal{C}}
\def\calM{\mathcal{M}}

\newtheorem{observation}{Observation}

\usepackage{natbib}
 \bibpunct[, ]{(}{)}{,}{a}{}{,}%
 \def\bibfont{\small}%
\EquationsNumberedThrough    

\TheoremsNumberedThrough     
\ECRepeatTheorems  %

\MANUSCRIPTNO{MOOR-0001-2024.00}
\def\cite#1{\citep{#1}}
\begin{document}


\RUNAUTHOR{}

\RUNTITLE{It Takes Two: A Peer-Prediction Solution for Blockchain Verifier's Dilemma}

\TITLE{It Takes Two: A Peer-Prediction Solution for Blockchain Verifier's Dilemma}



\ARTICLEAUTHORS{%

\AUTHOR{Zishuo Zhao}
\AFF{Department of Industrial and Enterprise Systems Engineering, University of Illinois Urbana-Champaign, Urbana, IL 61801, USA, \EMAIL{zishuoz2@illinois.edu}} 

\AUTHOR{Xi Chen}
\AFF{Leonard N. Stern School of Business, New York University, New York, NY 10012, USA, \EMAIL{xc13@stern.nyu.edu}
}
\AUTHOR{Yuan Zhou}
\AFF{Yau Mathematical Sciences Center \& Department of Mathematical Sciences, Tsinghua University, Beijing 100084, China, \EMAIL{yuan-zhou@tsinghua.edu.cn}
}
}


\ABSTRACT{The security of blockchain systems is fundamentally based on the decentralized consensus in which the majority of parties behave honestly, and the content verification process is essential to maintaining the robustness of blockchain systems. However, the phenomenon that a rational verifier may not have the incentive to honestly perform the costly verification, referred to as the Verifier's Dilemma, could incentivize lazy reporting and undermine the fundamental security of blockchain systems, particularly for verification-expensive \emph{decentralized AI} applications. 

In this paper, we initiate the research with the development of a \emph{Byzantine-robust peer prediction} framework towards the design of one-phase Bayesian truthful mechanisms for the decentralized verification games among multiple verifiers, incentivizing all verifiers to perform honest verification without access to the ground truth even in the presence of noisy observations, malicious players and inaccurate priors in the verification process, {proposing the compactness criteria that ensures such robustness guarantees}. With robust incentive guarantees and budget efficiency, our study provides a framework of incentive design for decentralized verification protocols that enhances the security and robustness of the blockchain, decentralized AI, and potentially other decentralized systems.}





\maketitle



\section{Introduction}


Blockchain, with prevailing examples as Bitcoin \citep{nakamoto2008bitcoin} and Ethereum \citep{buterin2014next}, is an emerging technology that maintains decentralized consensus via a distributed ledger that utilizes cryptographic techniques to achieve trust and security. {In recent years, the blockchain technology is  drawing wide interest in the operations research community (see, \citet{whitaker2020fractional,iyengar2022economics,manzoor2022blockchain,davydiuk2023crypto,doi:10.1287/opre.2022.0463,chen2025bayesian} and their references); on the other hand, it also has applications that empower traditional operations research studies, e.g., supply chain management (\citet{cole2019blockchain,keskin2023blockchain,cui2024supply}). Furthermore, following the current trend of artificial intelligence (AI), a frontier topic of \emph{decentralized AI} \cite{wang2024sok} occurs with the motivation to leverage blockchain technologies for securing training and inference procedures of machine learning (ML) computation to ensure credibility and accountability of AI models, and has been drawing interest in both blockchain and AI communities (see, e.g., \citet{chen2024zkml,conway2024opml,zhao2024proof}). 

In the meantime, from a game-theoretic perspective, a well-designed incentive mechanism is crucial to motivate self-interested players to behave honestly in a decentralized ecosystem, and a line of recent studies (see, \citet{roughgarden2021transaction,hansjoerg2022profitability,chen2024game, chen2025bayesian}, etc.) has formed an emerging research field of \emph{blockchain mechanism design} that investigates the design and analysis of on-chain reward/penalty\footnote{In the rest of this paper, we use terms ``reward'' and ``penalty'' interchangeably: a penalty can be regarded as a negative reward and vice versa.} mechanisms to incentivize honest behavior on blockchain platforms, including decentralized AI applications.}

{For concrete understanding of blockchain incentive mechanisms, let us look into how decentralized consensus is maintained.} In a blockchain system, the ``chain'' is essentially a linked list of ``blocks'' growing with time, where each block stores a piece of data (aka. \emph{transactions}). Every player stores a copy of the chain, and players who propose new blocks are supposed to simultaneously \emph{validate} previous blocks.
To achieve consensus in such decentralized systems, 
Bitcoin adopts the Proof-of-Work (PoW) that requires ``miners'' to expend significant computational effort (and energy) to earn access to blocks. 
Alternatively, Ethereum uses Proof-of-Stake (PoS) that {requires validators to stake tokens for participation, and} selects validators with probabilities in proportion to their staked tokens \cite{eip3675}. 
{Furthermore, recent studies also propose Proof-of-Learning (PoL) to replace the PoW task with AI model training \cite{jia2021proof,zhao2024proof}. \added{On the one hand, the PoL protocol serves as a Proof-of-Useful-Work (PoUW) that addresses the energy and sustainability issues of traditional PoW protocols \citep{vranken2017sustainability,stoll2019carbon} and the security and centralization issues of PoS \citep{bagaria2022proof}; on the other hand,} the verification of PoL in turn serves as the certificate that the model is trained honestly, realizing the motivation of decentralized AI \added{to prevent adversarial attacks on model training \cite{BytedanceAttack} and address \emph{AI safety} concerns \cite{bengio2024managing} in the current times}.}


\begin{figure}[tb]
    \centering
    \includegraphics[width = 0.75\textwidth]{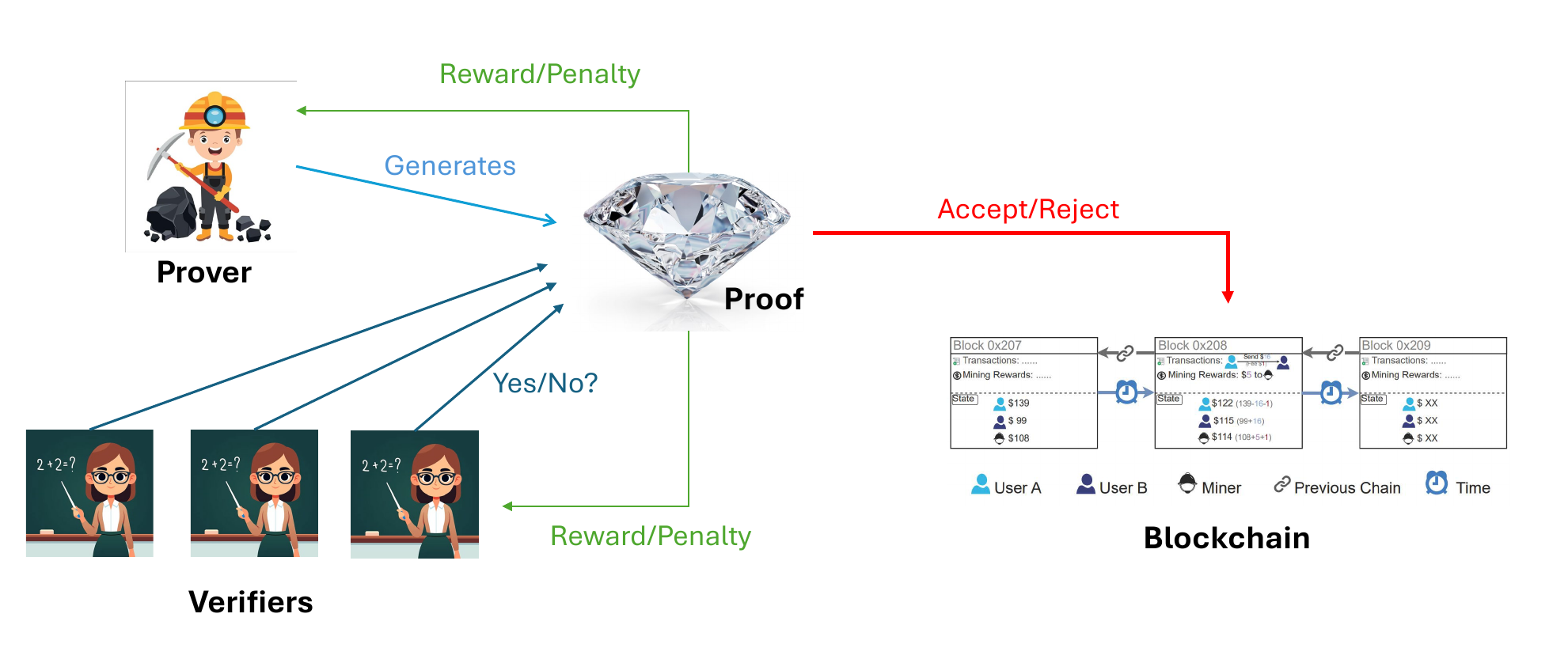}
    \caption{Illustration of decentralized verification games.}
    \label{fig:pos}
\end{figure}

While such mechanisms are motivated to incentivize the  prover to behave honestly, they can inadvertently introduce strategic concerns for validators, {e.g., rational validators may act \emph{lazily} or \emph{maliciously} in the validation process.}
A notable phenomenon is the Verifier's Dilemma \citep{luu2015demystifying,smuseva2022verifier,smuseva2025verifier}, {showing that rational verifiers do not have the incentive to act honestly} when a proposed block (or ``proof'') is honest with overwhelming probability and verification incurs nontrivial computational costs. While the Verifier's Dilemma does not appear to seriously undermine the security of Bitcoin or Ethereum in practice, it would become a prominent challenge in decentralized AI applications due to the heavy computational costs of ML verification \cite{conway2024opml,zhao2024proof}. The Verifier's Dilemma is described as follows:
{If a mechanism could incentivize provers to behave honestly, then no (rational) provers would cheat;
if no prover would cheat and the verification has a non-zero computational cost, then the verifier's optimal strategy is to lazily accept the proof without actual verification;
when verifiers become lazy, the incentive guarantees for provers no longer hold. }
Formally, the Verifier's Dilemma can be formulated as the following theorem. The proof is deferred to Appendix~\ref{app:veri:dil}. 

\begin{theorem}[Verifier's Dilemma]
\label{thm:verifier:dilemma}
In a verification game in which 

\begin{itemize}
\item A verifier's report is binary, e.g. ``Success'' or ``Fail'';
\item The verification result of an honest proof is always ``Success'';
\item Honest verification has a strictly positive cost, 
\end{itemize}
It is impossible to design an incentive mechanism realizing a pure-strategy Nash equilibrium such that the prover and verifier(s) simultaneously act honestly.\footnote{Although the Verifier’s Dilemma is sometimes understood as the tendency of verifiers to act \emph{lazily}, our study is aimed for a general purpose to design robust mechanisms incentivizing \emph{honest} verification---i.e., verifiers are incentivized not only to avoid laziness but also to refrain from acting maliciously.}
\end{theorem}

The Verifier's Dilemma arises in settings where the fraction of cheating provers tends to zero, making it optimal for verifiers to adopt lazy strategies. In such scenarios, any bounded reward for catching a cheat cannot, in expectation, offset the cost of verification. Hence, a traditional reward-based approach can only lower, but not eliminate, the rate of cheating in the pool of proofs. {This issue becomes particularly prominent in applications involving costly verification---particularly for 
decentralized AI applications. For example, \citet{conway2024opml} propose the opML (Optimistic Machine Learning), \added{a more straightforward protocol than PoL} for decentralized trustworthy ML computation performed by at least two parties, in which one \emph{prover} essentially runs the computation and one or more \emph{verifiers} \added{simply} re-run the same procedure for verification. \citet{conway2024opml} show that their mechanism achieves a mixed-strategy Nash equilibrium in which the prover has a $\frac{C}{R+L}$ probability to cheat, where $C$ is the verification cost and $R+L$ is essentially the penalty imposed to the verifier for failing to report a cheat plus the reward for successfully reporting one. In real-world blockchain systems, penalties are typically upper bounded by the required stake and excessive rewards may lead to inflation. As a result, higher verification costs exacerbate the dilemma: the system must either demand larger stakes, suffer severe inflation, or tolerate a higher fraction of dishonest proofs.



{Recent studies are actively working on addressing the challenge of Verifier's Dilemma, but they mostly depend on certain extents of trusted authorities.} A line of recent studies {attempts to bypass the binary-report assumption of Theorem~\ref{thm:verifier:dilemma} via introducing} ``attention challenges'' that contain extra information or deliberate errors, e.g., inserting deliberate objects (which can be valid or invalid) as so-called ``flags'' to incentivize verifiers to find and report, as in \citep{reynouard2024bar,teutsch2024scalable,luu2015demystifying,zhao2024proof,sheng2024proof,smuseva2025verifier},{ essentially bypassing the binary-report assumption in Theorem~\ref{thm:verifier:dilemma}}. {For example, \citet{zhao2024proof} propose a \added{incentive-secure PoL} protocol in which the verifiers are supposed to verify a random subset of the training process, in which the provers may use different designated random seeds as ``flags'' for the verifiers to report, \citet{sheng2024proof} design the flags as traces of transaction computation, and \citet{teutsch2024scalable} design the flags as deliberately invalid proofs.}  
However, in the scenario where the verifiers may also be strategic or even \emph{malicious}, the validity of verification results, particularly for ML computation, can also be costly to verify. Hence, in traditional decentralized ML verification protocols \citep{jia2021proof,zhao2024proof}, we generally need some credibility assumptions on verifiers. {Alternatively, other proposals invoke} additional phases of ``committee voting'' when disagreement occurs \cite{zhang2024proof,conway2024opml}, {use heuristic reputation-based designs \cite{xu2024decentralized}}, or resort to \emph{oracle}-like entities \cite{ezzat2022blockchain,nassirzadeh2024countchain}. {In these proposals, the committees, high-reputation parties, or oracles are regarded as trusted authorities and act} as \emph{proxies of ground truths}, {and these kinds of trusted authorities both lack theoretically guaranteed credibility (beyond heuristics) and undermine decentralization.}
As long as we want to design a fully decentralized system with no trusted authorities, the {ground truth} of proofs' validity may be inaccessible and payments can only be decided by consensus among the voting parties. {For reference, we defer detailed discussion on existing designs of decentralized AI protocols to Appendix~\ref{app:DeAI}. } 

In the {operations literature}, the technique of \emph{peer prediction} 
\citep{miller2005eliciting} refers to a wide scope of incentive mechanisms to elicit honest information {without access to ground truth}, which is widely adopted in the applications of dataset acquisition \citep{chen2020truthful}, peer grading \citep{dasgupta2013crowdsourced}, {and also recent blockchain applications \cite{cai2020truth,wang2023uncertainty}}. 
{A general paradigm of peer prediction is to {ask multiple players the same question (or overlapping question sets) and} reward each player based on the comparison between her report $Z_i$ and other players' reports $\Z_{-i}$ according to a subtly designed \emph{scoring rule}. 
As a toy example, in a 2-player simple-agreement scoring rule, the two players receive $+1$ when their reports agree, and receive $-1$ otherwise. In this case, the scoring rule $R_i(Z_1, Z_{2})$ is shown as Table~\ref{tbl:scoring}}. 

\begin{table}[tb]
\centering
\begin{small}
\begin{tabular}{|c|c|c|c|c|}
\hline
& $Z_2=0$ & $Z_2=F_1$ & $Z_2=F_2$ & $Z_2=1$ \\ \hline
$Z_1=0$ & $(1,1)$ & $(-1,-1)$ & $(-1,-1)$ & $(-1,-1)$ \\ \hline
$Z_1=F_1$ & $(-1,-1)$ & $(1,1)$ & $(-1,-1)$ & $(-1,-1)$ \\ \hline
$Z_1=F_2$ & $(-1,-1)$ & $(-1,-1)$ & $(1,1)$ & $(-1,-1)$ \\ \hline
$Z_1=1$ & $(-1,-1)$ & $(-1,-1)$ & $(-1,-1)$ & $(1,1)$ \\ \hline
\end{tabular}
\end{small}
\caption{The Simple-Agreement Scoring Rule}
\label{tbl:scoring}
\vspace{-1em}
\end{table} 

{Whereas the simple-agreement scoring rule may not theoretically incentivize truthful reporting in all scenarios, the general purpose of peer prediction studies  (including our research) is to design refined scoring rules that secure such incentive guarantees (for example, Table~\ref{table:scoring:rule:num} in Appendix~\ref{sec:exp}.)} {Nevertheless, whereas existing 
peer prediction mechanisms are designed to elicit truthful reports in the absence of ground truth, the following challenges occur in our setting {for blockchain and particularly decentralized AI applications}:}

\begin{itemize}
    \item Costly observation: Most traditional peer prediction mechanisms are designed to elicit truthful \emph{reporting} {without considering observation costs}, but in our setting we need to incentivize the verifiers to make costly computational efforts to verify on-chain contents, \added{particularly for decentralized AI applications such as opML and PoL in which the verification processes are computationally intensive (as discussed in Appendix~\ref{app:DeAI})}. 
    \item Robustness: Most traditional peer prediction mechanisms have strong assumptions that may not apply in the decentralized setting of blockchain ecosystems, particularly when players are anonymous and may be adversarial.  
    {Particularly, we need \textbf{\emph{permutation-proofness}} to disincentivize malicious reporting,  \textbf{\emph{Byzantine robustness}} against adversarial peer verifiers, and \textbf{\emph{distributional robustness}} against dishonest proofs.}
\end{itemize}

A widely adopted paradigm in literature is using mutual-information-based scoring rules \cite{kong2019information,zheng2024truthful}. While these types of scoring rules {provide \emph{permutation-proofness} and} are convenient for usage in scenarios with known prior, {they do not explicitly consider costly observation, and their Bayesian Nash equilibria  do not ensure the \emph{Byzantine robustness} if a small fraction of peers are malicious}. Furthermore, they also have a gap from resolving the Verifier's Dilemma due to {the lack of \emph{distributional robustness}}, as the logarithm-based scoring rules are sensitive to low probabilities. In the practice of blockchain systems, as the Verifier's Dilemma leads to an arbitrarily low cheating probability $\eps$, its empirical value becomes difficult to estimate and that sensitivity will severely undermine the robustness of the reward mechanism.

In contrast to most peer prediction mechanisms that guarantee Bayesian Nash equilibria requiring a known prior, a recent work \citep{kong2023dominantly} designs a determinant-based mutual information (DMI) mechanism without the need of prior information. 
However, it does not satisfy \emph{permutation-proofness}:
a verifier who systematically flips all her reports can still obtain optimal rewards, 
rendering it inapplicable for blockchain verification.
Conceptually, the DMI mechanism is motivated to elicit \emph{informative} (non-lazy) feedback rather than \emph{trustworthy} (non-lazy and also non-adversarial) ones. 
As the study of \citet{kong2019information} shows that {no peer prediction mechanism can satisfy prior-free and permutation-proof properties simultaneously},  the requirement of approximate prior knowledge remains necessary in the application of decentralized verification games.
Hence, assuming approximate prior knowledge is theoretically justified  in addressing the Verifier’s Dilemma in blockchain applications.}

\subsection{Our Contribution} 
\label{subsec:contribution}

In this research, we develop a theoretical framework with modeling of \emph{decentralized verification game (DVG)}, and initiate the study that combines the ideas of \emph{flags} and peer prediction into our proposed mechanism, named capture-the-flag peer prediction (CTF-PP), which only needs one phase in its procedure, and {incentivizes honest verifying and reporting via} simultaneously satisfying the following properties {that explicitly consider \emph{observation costs}}:

\begin{itemize}
    \item Interim \emph{unique} incentive compatibility (interim UniIC): A verifier, after performing the verification, maximizes her expected utility when she reports honestly. Furthermore, if she reports a different type from her observation, her expected utility is non-positive.
    \item Interim individual rationality (interim IR): A verifier, after performing the verification, gets a non-negative expected net utility, when she acts honestly.
    \item Interim no-free-lunch (interim NFL): A verifier cannot get a positive expected utility via any \emph{uninformed strategy} \cite{shnayder2016informed}, i.e., without doing the verification. 
\end{itemize}

Combining all the desired properties, we characterize the notion of \emph{incentive alignment} ($\delta$-IA) as a general guideline for peer prediction mechanisms in decentralized verification games, which depicts the property that any pure strategy gains a positive interim utility if \emph{and only if} it is honest, with a margin of $\delta$ (details in Section~\ref{subsec:ia})  {aimed for Byzantine robustness and distributional robustness}. With this stronger incentive guarantee, we can ensure that the peer prediction mechanism works as desired for the tricky setting of decentralized environments in blockchains, with an additional guarantee to \emph{disincentivize free-riding behavior} in blockchain systems, {particularly reinforcing economic foundations of decentralized AI ecosystems in which verification can be costly}. 

{We show the comparison of our design to existing peer prediction mechanisms in Table~\ref{tbl:comparison}. Beside theoretical derivations, we also perform extensive numerical experiments to show the effectiveness of our design in comparison with existing peer prediction mechanisms in different scenarios. (Section~\ref{sec:numerical} and Appendix~\ref{sec:exp}-\ref{app:exp2})}

\begin{table}[tb]
\centering
\begin{small}
\begin{tabular}{|c|c|c|c|c|c|}
\hline
& Prior & Observ. Cost & Perm. Proof & Byzan. Robust\tablefootnote{The maximum number of malicious verifiers that can be tolerated, $n$ denoting the total number of verifiers.} & Distr. Robust\tablefootnote{The maximum noise in prior distributions that can be tolerated, in the sense of total variation (TV) distance.{[Move to caption]}} \\ \hline
Log-Based\tablefootnote{Including family of Shannon entropy-based peer prediction mechanisms that use logarithm-based scoring rules, e.g., \citet{kong2019information,zheng2024truthful}.}& Needed & $\times$ & $\checkmark$ & $\times$ & $\times$ \\ \hline
DMI\tablefootnote{The mechanism proposed by \citet{kong2023dominantly}.}& Free & $\times$ & $\times$ & $n-1$ & $1$ \\ \hline
Ours& Only Approximate & $\checkmark$ & $\checkmark$ & $\Theta(n)$ & $\Theta(1)$ \\ \hline
\end{tabular}
\end{small}
\caption{Comparison of Peer Prediction Mechanisms for DVG.}
\label{tbl:comparison}
\vspace{-1em}
\end{table}

{Our technical contributions can be summarized as follows:}

\begin{enumerate}

\item 
We formulated {the desired incentive guarantees ($\delta$-IA) of $2$-verifier decentralized verification games (DVG) with a linear program (LP), and then illustrated the general feasibility of the linear program via a generalization of the Cremer-McLean mechanism \cite{cremer1988full}. Furthermore, we extend our methodology for general $n$-verifier DVG, showing a basic solution for DVG that incentivizes honest verification and reporting, considering \emph{observation costs} and satisfying \emph{permutation proofness} (Section~\ref{section:2v}).}
\item {More crucially, we discuss the} \emph{Byzantine robustness} properties and develop a general guideline of  the \emph{compactness} criteria that \emph{wide incentive margins and relatively low rewards/penalties} realize good Byzantine robustness against malicious verifiers, {and design an extended LP that additionally optimizes the compactness and achieves Byzantine robustness against an $\eps=O(1)$ fraction of malicious verifiers.} {We also show that the compactness criteria simultaneously realizes budget efficiency as such robustness guarantees only require an additional budget of $O(\eps)$.} (See in Section~\ref{sec:robustness})

\item Then, we observe the \emph{Byzantine reduction} principle {showing that inaccurate beliefs and priors can be statistically reduced to existence of malicious players via the \emph{coupling argument}}, and leverage this principle to connect the {\emph{distributional robustness}} against inaccurate priors/beliefs with the desiderata of the {Byzantine robustness}, showing the generality of our {compactness criteria} for robust peer prediction mechanisms. (See in Section~\ref{sec:inacc}) {[Move something here]}
\end{enumerate}



\section{Background and Related Work}


Since the emergence of Bitcoin \citep{nakamoto2008bitcoin}, the concept of blockchain is inherently designed as an unalterable distributed ledger that maintains trustworthiness via decentralized consensus. The blockchain can be modeled as a growing linked list stored by decentralized nodes, in which each \emph{block} contains its contents that consists of \emph{transactions}, a hash reference of its previous block, and a certificate (e.g., PoW, PoS and etc.) that controls the access to the block. Conceptually, when a block producer, also called a \emph{miner}, would like to pack and propose new contents on the blockchain, she needs to attach the block to a previous block, and pay certain efforts to gain access to produce the block. When a miner attaches a new block to an existing block, she is supposed to have also \emph{verified} the validity of the previous block. This process also makes the previous block unalterable, because the new block would be stored and witnessed by all the nodes of the network.

Nevertheless, in real-world blockchain ecosystems, the {verifiers} may be economically rational or selfish. In this context, the Verifier's Dilemma occurs. For example, \citet{cao2023leveraging} propose an attack that leverages the Verifier's Dilemma to double spend in Bitcoin. Besides, the studies of \citet{alharby2020data,smuseva2022verifier} make extensive analyses on Ethereum and the results show that Ethereum verifiers are frequently incentivized not to verify the contents while they are supposed to, rendering the Ethereum protocol vulnerable.

That said, one may argue that in the original design of Bitcoin or Ethereum, the verification of a block has negligible costs as it only needs the miner to check if all the transactions and the Proof-of-Work (PoW) or Proof-of-Stake (PoS) is valid. Since invalid blocks can be detected easily, miners might practically decide to behave honestly even if it is (slightly) irrational. Nevertheless, the development of the blockchain technology generalized the usage of blockchain system from an unalterable ledger of monetary transactions to a general decentralized platform that guarantees the integrity of diverse contents, e.g., smart contracts \citep{khan2021blockchain,ante2021smart,wang2019blockchain}, and furthermore, with the recent rapid development of AI technologies and the demands of trustworthy AI models, researchers are actively exploring to establish blockchain-based platforms that verify the computation of machine learning \cite{jia2021proof,nodehi2024game,chen2024zkml,conway2024opml,zhao2024proof}, which brings new motivations for blockchain studies as a novel paradigm of decentralized trustworthy AI.

Unlike hash puzzles in the Bitcoin PoW, the verification of such complicated contents can be potentially costly. 
Particularly, in the context of ML verification, \citet{fang2023proof} show that efficient byzantine-secure verification of stochastic gradient descent (SGD) computation reduces to fundamentally hard open problems in deep learning theories. Even though the study of \citet{zhao2024proof} achieves substantially lower verification overheads via the relaxation to incentive-security, the verification protocol still needs to reproduce the training process of at least $\Theta(1)$ epochs which has non-negligible computational costs. Consequently, recent studies typically resort to weaker incentive properties for verification games. For example, the recent proposal of opML~\cite{conway2024opml}, a protocol that designs for trustworthy ML inference on blockchain, can only reach a mixed-strategy Nash equilibrium that a (small) constant fraction of provers and verifiers behave dishonestly, and this fraction scales up with the verification cost (as discussed in Introduction). 

Because the Verifier's Dilemma, unless suitably addressed, appears as a fundamental vulnerability in the incentive structure of blockchains that may severely undermine the reliability of blockchain systems, the studies of \citet{teutsch2024scalable,zhang2024proof} work on this issue via introducing deliberate invalid objects as \emph{attention challenges} that incentivize verification. Nevertheless, their protocols are multi-phased as they need additional dispute processes and are potentially restricted to particular applications. In our work, we are motivated to design a one-phase general-purpose {and oracle-free} solution to the Verifier's Dilemma with theoretical incentive guarantees, expecting to resolve the critical incentive issue in decentralized verification games in a reliable and  efficient paradigm.



\section{Basic Modeling of Decentralized Verification Games}
\label{sec:model}

To initiate the study, we first formulate the modeling of decentralized verification games (DVG). In a $n$-verifier DVG, there are $n$ homogeneous players (verifiers) $i=1,\cdots,n$ independently verifying an on-chain \emph{proof}, and we use the terms ``player'' and ``verifier'' interchangeably.

The proof has an underlying ground-truth type $\theta \in S$, which can be either ``Honest'' ($\theta=0$), ``Flag $j$'' ($\theta=F_j$, for $j=1,2,\cdots,m$), or ``Dishonest'' ($\theta=1$). We define $S_*=\{0,F_1,\cdots,F_m\}$ as the set of all non-dishonest types. The actual type $\theta$ is unknown to both the verifiers and the system.  While the observations $\{X_i\}$ can potentially be noisy, every verifier's observation, when they actively verify the proof, is \emph{i.i.d.} conditioned on $\theta$ with known distributions $\{P(X_i|\theta)\}$.\footnote{The symmetry/homogeneity among verifiers can be assumed both according to the fixed verification protocols and the anonymity of decentralized systems.} Since the system can insert flags to maintain a pre-set flag rate (as in \citet{teutsch2024scalable,zhao2024proof,smuseva2025verifier}, etc.) that robustly incentivizes verification when no cheater occurs, we have \emph{principal} prior probabilities $P(\theta=F_i)=p_{F_i}$ and $P(\theta=0)=p_0$ as publicly known information, with $\sum_{\theta\in S^*}P(\theta)=1$ and the principal probability of $\theta=1$ is zero. Throughout the paper, the term ``principal'' refers exclusively to this cheater-free scenario.

However, in reality, the prior distribution slightly deviates from the principal scenario as there is a small but unknown probability $\epsilon\in [0,\epsilon_0]$ that the proof is dishonest, i.e., $P(\theta=1)=\epsilon$, with a known upper bound $\epsilon_0$. 
We assume that the appearance of dishonest proofs may take up the probabilities of types in $S_*$ in an arbitrary way. Hence, for any $s\in S_*$ we have $P(\theta=s)=p_s-\epsilon_s$, in which $\sum_{s\in S_*}\eps_s = \eps$ but the exact values of $\{\epsilon_s\}$ remain unknown.

Similar to \cite{zhao2024proof}, we begin our discussion with the verification process in a \emph{lossy-channel} model as follows, and then study the general case (as shown in Theorem~\ref{thm:feasibility}). When verifier $i$ verifies the proof, the distribution of the observation $X_i$ is dependent on $\theta$ in this way: 

\begin{itemize}
    \item \textbf{\emph{Completeness}}: A honest proof is always observed as honest, i.e. $P(X_i=0|\theta=0) = 1$.
    \item \textbf{\emph{Probabilistic soundness}}: A dishonest proof can be observed as any type, but the probabilities are known to the public, and the probability of correct detection at least $\kappa>0$, i.e. $P(X_i=1|\theta=1) \ge \kappa$. 
    \item \textbf{\emph{Benign flags}}: A flag $F_j$ can be detected with known probability $\mu_j>0$ or missed and observed as honest, but will never be observed as dishonest or other flags, i.e. 
    $P(X_i=F_j|\theta=F_j) = \mu_j$ and $P(X_i=0|\theta=F_j) = 1-\mu_j$.
\end{itemize}

\textbf{Verification protocol.} 
For each verifier $i$, {she first needs to \emph{stake} a pre-specified amount $L$ of tokens to the system, and is informed that she will be rewarded based on a public \emph{scoring rule} denoted as $R_i(Z_i, \Z_{-i})$, in which $Z_i$ is her report, $\Z_{-i}$ is the collection of other verifiers' reports, and the maximum possible penalty cannot exceed the staked amount, i.e., $R_i(Z_i, \Z_{-i})\ge -L$.  Then, the verifier makes a decision to follow one of the following strategies, or any mixture between them:}

\begin{enumerate}
\item Informed (active) strategy: Actively verifies the proof and gets the observation, which gains her access to $X_i$ but incurs a publicly-known cost $c(X_i)\ge 0$ which can depend on $X_i$.
\item Uninformed (lazy) strategy: Does not verify and has no access to $X_i$. For convenience of expression, we can denote $X_i=\bot$ in this case, and $c(\bot)=0$.
\end{enumerate}

Hence, verifier $i$'s Bayesian belief $\mathcal{B}$ on $\Z_{-i}$ is the conditional distribution of $P(\X_{-i}|X_i)$ for the informed strategy, or the marginal distribution $P(\X_{-i}|\bot)=P(\X_{-i})$ for the uninformed strategy. Here, $i$'s belief of the cheating probability can be an arbitrary $\eps^{(i)}\in [0,\epsilon_0]$ that can be different from the actual $\eps$, with arbitrary $\{\eps_s^{(i)}\}$ such that $\sum_{s\in S_*}\eps_s^{(i)} = \eps^{(i)}$, and we desire to design a mechanism that uniformly satisfies the incentive guarantees for arbitrary $\{\eps^{(i)}_s\}$. In this context, we define the \emph{principal belief} as the Bayesian belief given $\forall \eps^{(i)}_s=0$, i.e., $\eps_0=0$.

Then, verifier $i$ reports a $Z_i$ that maximizes $\E_{\tilde{\Z}_{-i}\sim \mathcal{B}}[R_i(Z_i,\tilde{\Z}_{-i})]$ in which $\mathcal{B}$ is her belief of $\Z_{-i}$, and claims that $Z_i$ is her observation. After each verifier $i$ independently reports $Z_i$ without seeing $\Z_{-i}$\footnote{This can be implemented via a cryptographic commitment scheme.}, the system has the information of $Z_1, \cdots,Z_n$, but not $\theta$, and rewards each prover $i$ according to the scoring rule $R_i(Z_i, \Z_{-i})$. {If $R_i(Z_i, \Z_{-i})<0$, the penalty will be deducted from her staked tokens.} Then, verifier $i$'s net utility is $R_i(Z_i, \Z_{-i})-c(X_i)$. 

\textbf{Formal characterization of strategies and utilities.} For the action of any verifier $i$, she may first decide to be active (informed) or lazy (uninformed). If she chooses to be lazy, then she can choose a distribution $D\in \Delta(S)$, in which $\Delta(S)$ is the set of all convex combinations of elements in $S$, and report $Z_i\sim D$. If she chooses to be active, she may observe $X_i$ and report according to a respective distribution corresponding to each $X_i\in S$, so any informed strategy can be characterized by a mapping $D(\cdot):S\to \Delta(S)$. Furthermore, the verifier can also randomly decide to be active or lazy. Hence, we formally characterize the verifiers' strategies as:

\begin{definition}
    The strategy space of any verifier can be characterized as 
    $$\Omega = \Omega(S) = \Delta(\{\Delta(S), \Delta(S)^S\}),$$
    and we denote the strategy of verifier $i$ as $s_i$. 
    
    For example, $s_i\in \Delta(S)$ if $i$ chooses an uninformed strategy, and $s_i\in \Delta(S)^S$ if $i$ chooses an informed strategy.
    Otherwise, if $s_i$ is a random choice between informed and uninformed strategies, we can represent $s_i$ with the 3-tuple $(\lambda, \mu, \alpha)$, denoted as 
    $$s_i\triangleq (\lambda,\mu,\alpha),$$ 
    in which $\lambda\in \Delta(S), \mu\in \Delta(S)^S, \alpha \in (0,1)$ and
    $s_i = \alpha\cdot \lambda + (1-\alpha)\cdot \mu.$
\end{definition}

For any verifier, she would maximize her expected utility based on her \emph{belief} on the reports of other verifiers. The belief profile of verifier $i$ can be characterized as $\mathbb{B}_i: (S\cup \{\bot\}) \to \Delta(S^{n-1})$ which maps her observation to a joint distribution of other verifiers' reports. For \emph{Bayesian} verifiers, they always set their belief as $\mathbb{B}_i(X_i)=P(\X_{-i}|X_i)$. Then, with regard to belief $\mathbb{B}_i$, we can define the interim utility $u_{i}(s_i;\mathbb{B}_i)$ as:
\begin{small}
\begin{align}
    u_{i}(s_i;\mathbb{B}_i) =
    \begin{cases}
    \E_{Z_i\sim s_i,\Z_{-i}\sim \mathbb{B}_i(\bot)}[R_i(Z_i,\Z_{-i})],\quad & s_i \in \Delta(S);\\
    \E_{X_i\sim P(X_i)}[\E_{Z_i\sim s_i(X_i), Z_{-i}\sim \mathbb{B}_i(X_i)} [R_i(Z_i,\Z_{-i})-c(X_i)]],\quad & s_i \in \Delta(S)^S;\\
    \alpha\cdot u_i(\lambda;\mathbb{B}_i) + (1-\alpha)\cdot u_i(\mu;\mathbb{B}_i), \quad &s_i \triangleq (\lambda,\mu,\alpha).
    \end{cases}
\end{align}
\end{small}
In most parts of this paper (except for the discussion of Byzantine robustness and Appendix~\ref{app:strong:scp}), we always assume that all verifiers have principal Bayesian beliefs $\mathbb{B}_i(X_i)=P(\X_{-i}|X_i)$ assuming $\eps=0$ (we justify using $\eps=0$ in place of small unknown $\eps>0$ in the robustness analysis of Section~\ref{sec:inacc}). Hence, we simplify the notation as $u_i(s_i)$. We call a strategy \emph{honest} or \emph{truthful} if and only if $s_i=I$ is (induced by) the identity map on $S$, i.e. $s\in \Delta(S)^S$ and $s(X_i)\equiv X_i$.

\textbf{Pure \& mixed strategies.} In the strategy space $\Omega$, we define the subset $\Omega_d = S \cup S^S$ as the space of \emph{pure} strategies, in which the verifier deterministically decides to be informed or uninformed, and reacts deterministically to her observation.
From the linearity of the utility function, we can see that the optimal utility is always realized by a pure strategy for any fixed belief, and hence we mainly consider pure strategies throughout this paper.

\def\eps{\epsilon}
\def\P0{\tilde{P}}

\subsection{IR and UniIC Constraints for Informed Verifiers}
\label{subsec:iric}

The IR constraint requires that given the verifier $i$ observes $X_i$, truthfully reporting it gains her an expected reward no less than $c(X_i)$. Since $X_i$ and $X_{-i}$ are independent conditioned on $\theta$, define $r_{X_i}(Z_i)$ as verifier $i$'s expected reward of reporting $Z_i$ conditioned on observing $X_i$, then $r_{X_i}(Z_i)$ can be computed as
\begin{small}
\begin{align}
    r_{X_i}(Z_i) &= \sum_{\X_{-i}\in S} R_i(Z_i,\X_{-i}) P(\X_{-i} | X_i) \\
    &= \sum_{\X_{-i}\in S} R_i(Z_i,\X_{-i}) \frac{P(X_i,\X_{-i})}{P(X_i)} \\
    &= \sum_{\X_{-i}\in S} R_i(Z_i,\X_{-i}) \frac{\sum_{\theta\in S} P(\theta) P(X_i|\theta) P(\X_{-i}|\theta) }{\sum_{\theta\in S} P(\theta) P(X_i|\theta)}.
\end{align}
\end{small}

While the probabilities are dependent on $\epsilon$, as long as $\epsilon$ is small enough, for $X_i\in S_*$, the $r_{X_i}$ is a continuous function w.r.t. $\epsilon$, so the IR constraints on $X_i\in S_*$ can be implied by
\begin{align}\label{eqn:ir:1}
    \sum_{\X_{-i}\in S} R_i(X_i,\X_{-i}) \frac{\sum_{\theta\in S} \P0(\theta) \P0(X_i|\theta) \P0(\X_{-i}|\theta) }{\sum_{\theta\in S} \P0(\theta) \P0(X_i|\theta)}
    \ge c(X_i) + \delta, \qquad \forall X_i\in S_*,
\end{align}
in which $\tilde{P}$ denotes the \emph{principal} probabilities assuming $\epsilon=0$, and the margin $\delta$ is introduced to ensure the incentive guarantees even if the actual priors slightly deviate from the principal priors, so that for any $\delta>0$, the constraints hold robustly for some $\epsilon_0 >0$. The condition is also necessary when $\delta=0$. In the rest of this paper, we say a condition is ``sufficient and almost necessary'' when it is sufficient with a $\epsilon_0>0$ depending on $\delta$, and when we set $\delta=0$, it becomes a necessary condition. We defer the rigorous justification of the ``sufficiency'' and quantitative analysis on the relations between $\delta$ and $\eps_0$ to Section~\ref{sec:inacc}.

For the case of $X_i=1$, from the lossy-channel model, we know that $\theta=1$. Therefore, we have
\begin{equation}\label{eqn:ir:2}
    r_1(1) = \sum_{X_{-i}\in S} R_i(1, \X_{-i}) P(\X_{-i}|\theta=1)
    \ge c(1)+\delta.
\end{equation}
So Eqs.~(\ref{eqn:ir:1}-\ref{eqn:ir:2}) are sufficient and almost necessary conditions that a CTF-PP mechanism is IR.

For the IC constraint, we need and only need $r_{X_i}(X_i) = \max_{Z_i\in S}\{ r_{X_i}(X_i)\}$. With similar arguments, we can also develop sufficient and almost necessary conditions that a CTF-PP mechanism is IC. Actually, given that the IR is satisfied we can define a stronger notion of \emph{Uniquely-IC} as follows:

\begin{itemize}
    \item Uniquely IC (UniIC): In 
addition to the IC requirement, given all other verifiers act honestly and a verifier actively performed the verification, then she gets a negative expected utility when she reports any type different from her observation.
\end{itemize}

Besides conventional IC notions, the UniIC requirement additionally rules out the possibility that a dishonest verifier cheats the system without losing money. Assuming that the IR constraints are already satisfied, the UniIC constraints can be formulated as:
\begin{equation}
    r_{X_i}(Z_i) \le c(X_i) - \delta, \qquad \forall X_i\in S, \:Z_i\ne X_i.
\end{equation}

\subsection{NFL Constraints for Uninformed Verifiers}
\label{subsec:nfl}

We assume verifiers other than $i$ are honest, i.e. they all decide on the informed strategy and $\Z_{-i}=\X_{-i}$. If verifier $i$ performs the uninformed strategy, she has no information on $\theta$ and her strategy can only be reporting any type in $S=\{0, F_1, \cdots, F_m, 1\}$, or any convex combination of them. Hence, $i$'s utility when she lazily reports $Z_i$ is denoted as:
\begin{equation}
    r_\bot(Z_i) = \sum_{\X_{-i}\in S} R_i(Z_i,\X_{-i}) P(\X_{-i}).
\end{equation}

From the NFL requirement and assuming small $\epsilon \le \epsilon_0$,  a sufficient and almost necessary condition is the following linear constraints
\begin{align}
    \sum_{\X_{-i}\in S} R_i(Z_i,\X_{-i}) P(\X_{-i}) \le -\delta,\qquad \forall  Z_i\in S.
\end{align}

\subsection{Incentive Alignment for Decentralized Verification Games}
\label{subsec:ia}

From the discussion in Section~\ref{subsec:iric}-\ref{subsec:nfl}, we would like to design a mechanism for the decentralized verification game that simultaneously satisfies IR, UniIC, and NFL constraints. Combining the derivations above, we can summarize the sufficient and almost necessary conditions that satisfy all constraints above. Hence, we define the notion of \emph{incentive alignment} ($\delta$-IA) as follows:

\begin{definition}[Incentive Alignment]
    A CTF-PP mechanism is $\delta$-incentive-aligned ($\delta$-IA) if and only if for any verifier $i$ and pure strategy $s_i\in \Omega_d$, 
    \begin{small}
    \begin{equation*}
    u_i(s_i) 
    \begin{cases}
    \ge \delta, &s_i=I; \\
    \le -\delta, &s_i\ne I.
    \end{cases}
    \end{equation*}
    \end{small}
    Equivalently,
    \begin{small}
    \begin{equation*}
        r_{X_i}(Z_i) - c(X_i) 
        \begin{cases}
        \ge \delta, &Z_i=X_i; \\
        \le -\delta, &Z_i\neq X_i.
        \end{cases}
    \end{equation*}
    \end{small}
\end{definition}

Here, the $\delta$-IA is a sufficient and almost necessary condition that IR, UniIC and NFL are simultaneously satisfied.


\section{Theoretical Guarantee for DVG: LP Modeling and Feasibility}
\label{section:2v}

In this section, we show a basic result on the existence of incentive aligned CTF-PP mechanisms for any 2-verifier DVG that satisfies mild conditions, {and then generalize our design to a general setting of $n$ verifiers.}

\subsection{The $2$-verifier Case}

Assume $\eps=0$, 
and define the \emph{principal} belief matrix $B: S^2\to \mathbb{R}$ as $B_{xy}=P(X_{-i}=y|X_i=x)$.\footnote{$B_1y$ can still be defined even if $
P(\theta=1)=\eps=0$, e.g. $\theta=1$ when a zero-measure set is drawn.}
Besides, we define $B_{\bot}$ as the blind-belief (row) vector as $B_{\bot y} = P(X_{-i}=y)$ that describes the belief of verifier $i$ when she does not verify the proof. Then, we can formulate the design of a $\delta$-IA CTF-PP mechanism as a linear programming (LP) problem.

We define decision variable as the scoring matrix $T: S^2\to \mathbb{R}$ with $T_{xy}=R_i(x,y)$, and denote 
\begin{equation}\label{eqn:lp:w}
    W = BT'.
\end{equation}
Then $W_{xy}=r_x(y)$, which is the expected reward verifier $i$ gets from the mechanism when she observes $x$ and reports $y$. The IR and UniIC conditions are equivalent to the following:
\begin{small}
\begin{align}\label{eqn:lp:1}
    W_{xx} &\ge c(x) + \delta,\qquad \forall x\in S; \\
    W_{xy} &\le c(x) - \delta,\qquad \forall x\in S,\quad y\in S-\{x\}.\label{eqn:lp:2}
\end{align}
\end{small}
Similarly, we denote 
\begin{equation}
    W_\bot=B_\bot T',
\end{equation}
then $W_{\bot y}=r_\bot(y)$ is the expected reward verifier $i$ gets when she does not verify and lazily reports $y$. Then the NFL conditions are equivalent to the following:
\begin{equation}
    W_{\bot} \le -\delta. \label{eqn:lp:3}
\end{equation}
Hence, we only need to find a feasible solution of the linear system (\ref{eqn:lp:w}-\ref{eqn:lp:3}), i.e., solve the following linear program:
\begin{small}
\begin{align*}
    LP_0:\quad minimize\quad & 0 \\
    s.t. \quad&\textup{(\ref{eqn:lp:w}-\ref{eqn:lp:3})}. \label{eqn:lp0}
\end{align*}
\end{small}
Here, {inspired by the Cremer-McLean mechanism \cite{cremer1988full}}, we propose our {basic} theorem that shows the feasibility of $LP_0$, with the proof deferred to Appendix~\ref{app:feasibility}:
\begin{theorem}[{Basic} Theorem]\label{thm:feasibility}
    If $B$ is invertible, and $ P(X_i=y|\theta=1) = 0$ for any $y\ne 1$ (i.e., a non-cheating proof is never observed as a cheat), then for any $\delta\ge 0$, we can find a $\delta$-IA  mechanism for the 2-verifier DVG as a feasible solution of $LP_0$.
    
    For some $\eps_0>0$, the mechanism is IR, NFL and UniIC for any $\eps\in [0,\eps_0]$.
\end{theorem}

Particularly, we show that our method always works for the DVG with the lossy-channel model as defined in Section~\ref{sec:model} with the following proposition. The proof is deferred to Appendix~\ref{app:invert}. 

\begin{proposition}\label{prop:invert}
    In the lossy-channel model defined in Section~\ref{sec:model}, the principal belief matrix $B$ is invertible.
\end{proposition}

%


\subsection{General $n$-Verifier Case}
\label{section:nv}

We have just shown that under mild assumptions there always exists an incentive-aligned mechanism for any 2-verifier DVG. In this part we invoke the 2-verifier mechanism as a building block and construct our mechanism for the general setting of $n$ verifiers.

\textbf{Vectorized notation.} In the 2-verifier game, the $(Z_i,Z_{j})$-th entry of the matrix $T$, denoted as $T_{Z_iZ_{j}}$, depicts the reward of verifier $i$ when she reports $Z_i$ while the other verifier $j$ reports $Z_j$. With a slight abuse of notation, if we regard each type in $S$ as a unit one-hot column vector in the corresponding dimension, we can get $T_{Z_iZ_j} = Z_i'TZ_j$. In the general case of $n$ verifiers, we use a pairwise-scoring mechanism that compares every verifier's report with the \emph{average} of other verifiers', which can be formulated as:
\begin{equation}
    R_i(Z_i,\Z_{-i}) = Z_i' T \overline{\Z_{-i}}.\label{eqn:linear:mechanism}
\end{equation}
Here, we denote
\begin{equation}
    \overline{\Z_{-i}} = \frac{1}{n-1}\sum_{j\ne i} Z_{j}.
\end{equation}

Then, we can show that the pairwise-scoring mechanism as described as Eq.~\eqref{eqn:linear:mechanism} has equivalent incentive structures as the 2-verifier mechanism characterized as $T$. Formally, we have:

\begin{theorem}\label{thm:generalization}
    If the scoring matrix $T$ satisfies the $\delta$-IA property for the 2-verifier DVG, then the scoring rule as Eq.~\eqref{eqn:linear:mechanism} also satisfies $\delta$-IA for the general $n$-verifier DVG.

    Furthermore, if the 2-verifier mechanism characterized as $T$ is IR, NFL and UniIC for any $\eps\in[0,\eps_0]$, then the $n$-verifier mechanism in Eq.~\eqref{eqn:linear:mechanism} also satisfies the same properties.
\end{theorem}

The proof of Theorem~\ref{thm:generalization} is deferred to Appendix~\ref{app:gen:proof}.

\section{Byzantine Robustness via Margin Optimization}
\label{sec:robustness}

In the context of (Bayesian) Nash equilibria, we aim to design mechanisms in which no agent may benefit from \emph{individual} deviations. In other words, we guarantee that each verifier is incentivized to be honest given that \emph{all} others are honest. However, in decentralized ecosystems like blockchains, this assumption may be too strong as there may exist \emph{malicious} players who would deliberately attack the system, i.e., trying to undermine the robustness of the system at the risk of losing their own utilities. Furthermore, just like the widely studied topic of blockchain \emph{transaction fee mechanisms} (See, e.g., \citet{roughgarden2020transaction,roughgarden2021transaction,shi,wu2023maximizing,chen2025bayesian}), blockchain players may also potentially collude with each other or create fake identities to increase their utilities. Because the blockchain consensus protocols, e.g., PoW or PoS, can inherently address the Sybil attack issue, in this study we mainly consider non-Sybil dishonest players who do not conduct Sybil attacks but may act adversarially otherwise.


Whereas it may be too strong to assume that all other players are individually rational, in the field of decentralized systems, the notion of \emph{Byzantine robustness} (See, e.g., \citet{yin2018byzantine,wu2023byzantine,chen2012robustness}), also called \emph{Byzantine fault tolerance} or \emph{Byzantine resilience}, is widely studied as a desired property that the system works robustly as expected even if a (small) portion of the system does not work correctly. In the works of \citet{schoenebeck2021information,wang2023uncertainty}, the existence of colluding players is also considered for peer prediction mechanisms. Particularly, \citet{schoenebeck2021information} consider the multi-task setting and tackle with it as a \emph{robust learning} problem, and \citet{wang2023uncertainty} focus on the specific \emph{leader election} problem \cite{gharehchopogh2014survey} for blockchain consensus. 
In another perspective, \citet{frongillo2017geometric} look into the scenario of peer prediction with inaccurate distributional knowledge, and develop a margin optimization methodology to maximize the tolerance to inaccurate posterior beliefs.

Inspired by these studies, we are motivated to design a general-purpose solution for decentralized consensus with stronger \emph{incentive alignment} guarantees, with an optimization framework of Byzantine robustness {via the the \emph{compactness} criteria (See in Section~\ref{subsec:br:criteria}.)} Furthermore, we {show the budget efficiency of our design in Section~\ref{subsec:budget}}, and will show in Section~\ref{sec:inacc}  the generality of our Byzantine robustness notion as it can also imply \emph{distributional robustness}.
Following the framework in the study of \citet{schoenebeck2021information}, the types of players can be generally classified into the following categories:

\begin{enumerate}
    \item Altruistic ($\calA$): Acting honestly without consideration of utilities;
    \item Selfish ($\calS$): Acting in the way that maximizes their own utilities;
    \item Colluding ($\calC$): Conducting collusions with other players (within $\calC$) to maximize their joint utility;
    \item Malicious ($\calM$): Acting in arbitrarily manners that may not optimize their utilities, without access of non-malicious players' information.
\end{enumerate}

While traditional game theory primarily focuses on the behavior of $\calA$ and $\calS$ players, $\calC$ and $\calM$ players typically fall outside the scope of its standard models. Therefore, we call $\calA,\calS$ players as \textbf{{benign}} and $\calC,\calM$ players as \textbf{{rogue}}. 

Intuitively, we would like to design the mechanism in the following way: as long as rogue verifiers only constitute a small portion, all four types of verifiers are incentivized to act honestly, even though malicious players may actually act in different manners at the cost of their own utilities. However, we still assume that malicious players cannot access non-malicious players' information (e.g. observations and reports) as the unauthorized access of non-malicious players' information should be prevented by the system design, and also breaks the basic model of Bayesian games.

\subsection{Characterization of Robust Incentive Properties}
\label{subsec:robust:incentive}

To ensure that arbitrary actions of malicious players do not affect the incentive guarantees of other players even in the worst case, we introduce the notion of \emph{robust incentive properties} describing the scenario in which the incentive properties holding uniformly for any possible realization of malicious players' actions. 

In this case, we can define $\phi$ as the \emph{environmental variable} that depicts the prior probabilities, the number of $\calA, \calS, \calC, \calM$ players, and the strategies of $\calM$ players. While players might not have the exact information of $\phi$, they do have the knowledge that $\phi$ lies in a set $\Phi$ of \emph{environmental assumptions}, e.g., the total number of $\calC, \calM$ players does not exceed a particular threshold.

The motivation of \emph{robust incentive properties} is to guarantee that the players will not regret their honest actions even if they learn the existence and actions of the malicious players \emph{ex-post}, so that these malicious behavior would not affect the incentive guarantees for the majority of non-malicious players. In other words, in a mechanism with robust incentive guarantees, the desired properties hold uniformly for any $\phi\in\Phi$, {similar to the framework of \emph{distributionally robust optimization} \cite{kuhn2025distributionallyrobustoptimization}.}

In this context, given different environment variables, the player $i$ would have different beliefs of the other players' reports, and we can characterize the belief profile of player $i$ as $\mathbb{B}_i:(S\cup\{\bot\})\times \Phi \to \Delta(S^{n-1})$, which maps the tuple of her observation and the environment variable to a joint distribution of other players' reports.
Different from the model in Section~\ref{sec:model}, the players do not have a prior distribution of $\phi\in\Phi$. Hence, similar to the characterization of partial distributional knowledge in \cite{zheng2021limits}, for any strategy $s_i$, player $i$ would actually have a \emph{belief interval} $[u^{\Phi-}_i(s_i;\B_i),u^{\Phi+}_i(s_i;\B_i)]$ of its utility among all possible $\phi$'s, formulated as
\begin{small}
\begin{align*}
    &u^{\Phi-}_i(s_i;\B_i) = \inf_{\phi\in\Phi}u^{\phi}(s_i;\B_i),\nonumber\\
    &u^{\Phi+}_i(s_i;\B_i) = \sup_{\phi\in\Phi}u^{\phi}(s_i;\B_i),
\end{align*}
\end{small}
in which
\begin{small}
\begin{align}
&u^{\phi}_i(s_i;\B_i)= 
\begin{cases}
    \E_{Z_i\sim s_i,\Z_{-i}\sim \mathbb{B}_i(\bot,\phi)}[R_i(Z_i,\Z_{-i})],\quad & s_i \in \Delta(S);\\
    \E_{X_i\sim P(X_i)}[\E_{Z_i\sim s_i(X_i), Z_{-i}\sim \mathbb{B}_i(X_i,\phi)} [R_i(Z_i,\Z_{-i})-c(X_i)]],\quad & s_i \in \Delta(S)^S;\\
    \alpha\cdot u_i(\lambda;\mathbb{B}_i) + (1-\alpha)\cdot u_i(\mu;\mathbb{B}_i), \quad &s_i \triangleq (\lambda,\mu,\alpha).
    \end{cases}
\end{align}
\end{small}

In this section, we also mainly consider \emph{Bayesian} players with $\B_i(X_i,\phi)=P(\X_{-i}|X_i,\phi)$, and denote $u^{\phi}_i(s_i)$ ,$u^{\Phi-}_i(s_i)$, $u^{\Phi+}_i(s_i)$ in this case for simplicity. Then with the general guideline that the incentive properties should uniformly hold for any $\phi\in\Phi$, we define the notion of robust utility maximization and robust IA as follows:

\begin{definition}[Robust Utility Maximization]
In a fixed mechanism, a strategy $s_i$ robustly maximizes player $i$'s utility w.r.t. environmental assumption $\Phi$, if and only if for any strategy $s_i'\in \Omega$,
\[
    u^\phi(s_i)\ge u^\phi(s_i'),\qquad\qquad\qquad \qquad\quad\forall \phi\in\Phi.
\]
\end{definition}

\begin{definition}[Robust Incentive Alignment]
A mechanism satisfies \emph{robust} $\delta$-IA w.r.t. environmental assumption $\Phi$, if and only if for any player $i$ and pure strategy $s_i\in \Omega_d$,
\begin{align*}
    u_i^{\Phi-}(s_i) &\ge \delta,& s_i=I;\\
    u_i^{\Phi+}(s_i) &\le -\delta,&s_i\ne I.
\end{align*}
Here, we denote $I$ as the honest informed strategy, i.e., $s_i \in \Delta(S)^s$ and $s_i(X_i)\equiv X_i$.

\end{definition}

From the notions above, we formally define the notion of $f(n)$-Byzantine-robustness ($f(n)$-BR) as follows: 

\begin{definition}[Byzantine Robustness]
\label{def:br}
For a DVG with $n$ verifiers, we call a mechanism $f(n)$-BR if and only if: as long as $\Phi$ constrains that the total number of rogue ($\calC$ and $\calM$) verifiers does not exceed $f(n)$, 
\begin{itemize}
    \item Each $\calA, \calS, \calC$ verifier robustly maximizes her interim utility via acting honestly with robust $0$-IA guarantees, assuming that other $\calA, \calS, \calC$ verifiers act honestly.
    \item Each colluding party in $\calC$ robustly maximizes their total interim utility via acting honestly, assuming that all $\calA, \calS$ verifiers and other colluding parties in $\calC$ act honestly.
    \item Each $\calM$ verifier would robustly maximize their interim utilities with robust $0$-IA guarantees if she acted honestly, even though she may actually act otherwise, assuming that all $\calA, \calS, \calC$ verifiers act honestly.
\end{itemize}
\end{definition}

In the rest of this section, we show that  under mild assumptions, the design in Section~\ref{section:nv}, as long as  the scoring matrix $T$ comes from a ``good'' solution of the linear system Eqs.~(\ref{eqn:lp:1}-\ref{eqn:lp:3}), is $\Theta(n)$-BR, i.e., resilient against a constant portion of rogue verifiers.

\subsection{Bang for the Buck: Compactness Criteria for Byzantine Robustness}
\label{subsec:br:criteria}

When we look at the pairwise-scoring mechanism Eq.~\eqref{eqn:linear:mechanism}, the reward of each verifier $i$ is essentially based on the comparison of her report $Z_i$ and the \emph{average} of other verifiers' reports  $\overline{\Z_{-i}}$. Intuitively, if only a small portion of other verifiers may act dishonestly, since their contribution to $\overline{\Z_{-i}}$ is not significant, the actual expectation of $\overline{\Z_{-i}}$ conditioned on $X_i$ would not deviate significantly from $\E\big[\overline{\X_{-i}}\big|X_i\big]$, and the $\delta$ margin in our design would make the reward matrix of $i$ still satisfy \emph{incentive alignment} properties even with a slightly perturbed posterior distribution of $\overline{\Z_{-i}}$.

For simplicity of discussion, in the family of rogue verifiers, we first only consider \emph{simple malicious} ones who could not create fake identities but may act unpredictably and report in any strategy, as in the \emph{canonical Byzantine setting} defined in Definition~\ref{def:canon:byzantine} below. In later sections, we will show that (certain types of) collusions can also be reduced to the canonical Byzantine setting (details in Section~\ref{subsec:collusion:reduction}). For Sybil attacks, while no voting-based protocols can effectively prevent them if the attacker has unlimited resources (e.g., 51\% attack \citep{raju2022overview}), each Sybil identity can also be regarded as a (new) malicious agent and, as long as they only make up a small portion of all verifiers, our framework of Byzantine robustness can prevent them from harming the incentive structure of other verifiers. 
Furthermore, since gaining additional voting power in PoW or PoS protocols has additional costs, we can show that as long as the total resources (e.g., computing power for PoW or stakes for PoS) of the verifier only makes up a small portion of the network, she could not gain a significant advantage via Sybil attacks compared to honest behavior. 

\begin{definition}[Canonical Byzantine Setting]
In a canonical Byzantine setting, each verifier acts in one of the following strategies:
\label{def:canon:byzantine}
\begin{itemize}
    \item No-Sybil selfish ($\calS_*$): Acting in a way that maximizes their own utilities, but unable to create fake identities.
    \item No-Sybil malicious ($\calM_*$): Reporting arbitrarily, but unable to create fake identities.
\end{itemize}
\end{definition}

From the pairwise-scoring mechanism, we can see that conditioned on the verifier $i$ observing $X_i\in S\cup \{\bot\}$, the expected utility of reporting $Z_i$ is
\begin{equation}
r_{X_i}(Z_i) - c(X_i) = \frac{1}{n-1}\sum_{j\ne i}\left(\sum_{Z_j\in S} P(Z_j|X_i) T_{Z_iZ_j} - c(X_i)\right).
\end{equation}
When verifier $j$ is honest, we see that $Z_j=X_j$ and $\sum_{X_j\in S} P(Z_j|X_i) T_{Z_iZ_j}-c(X_i) = (BT')_{X_iZ_i}-c(X_i) = W_{X_iZ_i}$, and the $\delta$-IA condition ensures that $W$'s diagonal entries are at least $+\delta$ and other entries are at most $-\delta$. If $j$ is dishonest, then her report $Z_j$ may deviate from $X_j$, resulting in a different $\sum_{X_j\in S} P(Z_j|X_i) T_{Z_iZ_j}$ and leading to a perturbation to $r_{X_i}(Z_i)$.

Intuitively, if the summation of all these perturbations is bounded below $\delta$, then $\{r_{X_i}(Z_i)-c(X_i)\}$ still has positive diagonal entries and negative non-diagonal entries, satisfying the robust incentive alignment property (with a smaller margin). On the other hand, we notice that $\sum_{X_j\in S} P(Z_j|X_i) T_{Z_iZ_j}$ is a convex combination of $\{T_{Z_iZ_{j}}:Z_j\in S\}$. If we upper bound the magnitude of the scoring rule, i.e.
\begin{equation}
    |T_{Z_iZ_j}| \le K, \quad \forall Z_i,Z_j \in S,
\end{equation}
then we can deduce that 
\begin{equation}
    \sum_{Z_j\in S} P(Z_j|X_i) T_{Z_iZ_j} \in [-K,K].
\end{equation}

Hence, each dishonest verifier $j$ can perturb the value of $r_{X_i}(Z_i)$ by at most $\frac{2K}{n-1}$, so a large incentive margin $\delta$ with a relatively small $K$ would achieve a good \emph{``bang for the buck''} for desired Byzantine-robust guarantees. In this sense, we define $(\delta,K)$-compactness as:

\begin{definition}[$(\delta,K)$-compactness]
For fixed observation costs $c(\cdot)$ and principal belief matrix $B$ (denoted as the $(c,B)$-environment), a pairwise scoring matrix $T$ is called $(\delta,K)$-compact if and only if its entries are bounded within $[-K,K]$ and the corresponding mechanism is $\delta$-IA.

 For convenience, we also call a mechanism (or a pairwise scoring matrix) $\frac{\delta}{K}$-compact if it is $(\delta,K)$-compact for some $(\delta,K)$.
\end{definition}

Then, we immediately derive the following lemma, {showing that $(\delta,K)$-compactness implies the Byzantine robustness against a $\Theta(\frac{\delta}{K})$ fraction of malicious players}:

\begin{lemma}\label{lem:byzantine}
    If a CTF-PP mechanism has a $(\delta,K)$-compact pairwise scoring matrix, then it is $\frac{\delta}{2K}(n-1)$-BR in the canonical Byzantine setting, as it is $0$-IA even in the presence of up to $\frac{\delta}{2K}(n-1)$ malicious players. 
\end{lemma}

In the following parts we focus on the construction of $(\delta,K)$-compact pairwise scoring matrices with optimized $\frac{\delta}{K}$.

\subsection{LP Modeling for Byzantine Robustness}

In Section~\ref{section:2v}, we formulated the $\delta$-IA condition for the 2-verifier DVG as the linear system of (\ref{eqn:lp:1}-\ref{eqn:lp:3}), and showed that the linear system is generally feasible, so that a desirable mechanism can be found via linear programming, and further showed that the LP solution generalizes to the $n$-verifier setting, so that our proposal is a general-purposed solution for the design of DVG mechanisms. 

Whereas the general paradigm of incentive requirements in peer prediction can be depicted with the linear system, the LP problem actually allows us to optimize an \emph{objective function}, which is not specified in previous parts. Considering the motivation of Byzantine robustness, from Lemma~\ref{lem:byzantine} we would like to construct a $(\delta,K)$-compact scoring matrix with a large $\frac{\delta}{K}$. Hence, an intuitive idea is to fix $\delta$ and minimize $K$. For fixed principal belief matrix $B$, prior distribution vector $B_\bot$
and observation costs $c(\cdot)$, we can formulate the LP problem $LP_1(B,B_\bot,c,\delta)$ with decision variable $\big(K\in\R, T\in \R^{S^2}\big)$ as:
\begin{small}
\begin{align}
&LP_1(B,B_\bot,c,\delta):\hspace{-8em}\nonumber\\
    \quad &minimize\hspace{-2em} & K \\
    &s.t. &|T| &\le K,\\
    &&(BT')_{xx}&\ge c(x)+\delta, \:\forall x\in S~ &\label{eqn:lp1:cond3}\\
    &&(BT')_{xy}&\le c(x)-\delta, \:\forall x\in S, \: y\in S-\{x\}\label{eqn:lp1:cond4}\\
    &&B_\bot T' &\le -\delta.\label{eqn:lp1:cond5}
\end{align}
\end{small}
In fact, denoting $K_*=(B,B_\bot,c,\delta)$ as the optimal objective value of $LP_1(B,B_\bot,c,\delta)$, we can show an upper bound on $K_*(B,B_\bot,c,\delta)$ as:

\begin{theorem}\label{thm:Nbound}
    Denote $c_{1}=\max_{x\in S}\{c(x)\}$ as the maximum observation cost, $p_1=\max\{B_\bot\}=\max_{x\in S}\{P(X_i=x)\}$ as the maximum prior probability of any observation, and $k=|S|=m+2$ as the number of types, then we have
    \begin{equation} \label{eqn:n:ub}
        K_*(B,B_\bot,c,\delta) \le \|B^{-1}\|_2\cdot \left(c_1\cdot g_1(k,p_1) + \delta\cdot g_2(k,p_1)\right),
    \end{equation}
    in which
    \begin{small}
    \begin{align}
        g_1(k,p_1) &= \sqrt{\Big(1+ (k-1)\frac{p_1}{1-p_1}\Big)\Big(1+\frac{1}{1-p_1}\Big)}=O\left(\frac{\sqrt{kp_1}}{1-p_1}\right),\\
        g_2(k,p_1) &= \sqrt{\Big(k+(2k-2)\frac{p_1}{1-p_1}\Big)\Big(k+\frac{2}{1-p_1}\Big)}=O\left(\max\left\{k\sqrt{\frac{p_1}{1-p_1}},\frac{\sqrt{kp_1}}{1-p_1}\right\}\right).
    \end{align}
    \end{small}
    are only dependent on $k,p_1$ but independent to $n$.

    Additionally, there exists a feasible solution satisfying \eqref{eqn:n:ub} that makes the equality hold in \eqref{eqn:lp1:cond3}.
\end{theorem}

The proof of Theorem~\ref{thm:Nbound} is deferred to Appendix~\ref{app:Nbound}. 

\subsection{Reduction of Colluding Players}
\label{subsec:collusion:reduction}

For the characterization of Byzantine players, it is intuitive that colluding behavior is within the scope of malicious behavior, and \emph{the resilience against malicious players should infer the resilience against colluding players.} While this proposition is true, the reduction is actually non-trivial.

From the classification of players, we only consider the \emph{external} effects and \emph{individual} incentives of malicious players, i.e., their existence does not disrupt the incentive alignment guarantees of \emph{other} players or benefit individual utilities. However, in the consideration of colluding players, we still need to prevent them from gaining \emph{total} utility via collusion, i.e., we also need to consider \emph{internal} effects of collusion which is not covered in previous discussion. Similar to the Side-Contract-Proofness (SCP) notion proposed by \citet{shi}, we define \emph{weak-SCP} in the scope of DVGs as follows:

\begin{definition}[weak-Side-Contract-Proofness (weak-SCP)]
    We further define $\calC_*$ players as:
    \begin{itemize}
        \item No-Sybil colluding ($\calC_*$): Conducting collusions with other players (within $\calC_*$) to maximize their joint utility, but unable to create fake identities.
    \end{itemize}
    We call a mechanism weak-Side-Contract-Proof (\emph{weak-SCP}) under some conditions, if and only if as long as these conditions hold, in any collusion party, the players robustly maximize their total interim utility w.r.t. their individual (non-shared) Bayesian beliefs ($P(X_{-i}|X_i)$) via acting honestly.
\end{definition}

We call this notion \emph{weak}-SCP because although players collude with each other, they still update their posterior beliefs only based on their own observations, not considering other colluders' observations as they might not ``fully trust each other''. In turn, we call the collusion-proofness against colluders who share their beliefs based on their aggregated observations \emph{strong-SCP}.
Nevertheless, there are additional challenges in the design of strong-SCP peer prediction mechanisms and we leave it to future work. The intuition behind such challenges is that \emph{a belief-sharing colluding party would be incentivized to report the same type even though they may have different observations.} On the bright side, from Proposition~\ref{prop:colred}, although belief-sharing colluders may benefit from collusions, they do not disturb the incentive guarantees of other players as long as they only constitute a small portion of all players. We defer detailed discussions on strong-SCP to Appendix~\ref{app:strong:scp}.

In actual cases, there may exist multiple colluding parties, but from the argument in the player classification, for any selfish player or colluding party that intend to maximize their total utility, other colluding players outside the party do not have access to their actions or observations, and only need to be considered w.r.t. their external effects. Hence, we can deduce that

\begin{proposition}
\label{prop:colred}
Colluding players can be regarded as malicious players from the perspective of players outside their colluding parties.
\end{proposition}

With Proposition~\ref{prop:colred}, we only need to consider the existence of one colluding party of collusion players, beside a (small) number of malicious players. Formally, we have the following theorem:

\begin{theorem}\label{thm:scp}
Assume that there are $n$ players, among which are $|\calM_*|$ no-sybil malicious players and a no-sybil colluding party of $|\calC_*|$ players. If the CTF-PP mechanism has a $(\delta,K)$-compact scoring matrix, then the mechanism is $0$-IA and weak-SCP as long as 
$$|\calM_*|+|\calC_*| \le \frac{\delta}{2K}(n-1).$$
\end{theorem}

The proof of Theorem~\ref{thm:scp} is deferred to Appendix~\ref{app:scp}.
From Theorem~\ref{thm:scp}, we show that in the weak-SCP notion, colluding players can also be reduced to malicious players w.r.t. the weak-SCP notion for the Byzantine-robustness results of Theorem~\ref{thm:Nbound}.

\subsection{{Budget and Cost of Robustness}}
\label{subsec:budget}

In real-world information elicitation and crowdsourcing applications, the principal may also aim to minimize the total budget while maintaining the desired level of information quality. Even within blockchain ecosystems, where tokens can be minted at will, excessive token issuance should be avoided to prevent inflation and the consequent devaluation of the cryptocurrency.

In the specific context of decentralized verification games, our goal is ideally to compensate verifiers exactly according to their (expected) verification costs—this serves as a natural lower bound on the total budget, provided that individual rationality (IR) constraints are satisfied. Moreover, to ensure robustness, we may impose an additional $\delta$-margin as specified by the $(\delta,K)$-compactness criterion. Under this requirement, the lower bound on the budget becomes the expected verification cost plus $\delta$, where the $\delta$ term can be interpreted as the cost of robustness.

For this analysis, we assume $\epsilon = 0$ and that all verifiers are honest. Let $\overline{c}$ denote the expected verification cost, and $\overline{r}$ the expected payment to a verifier, both taken over the distribution of ground-truth states $\theta$ and corresponding observations. We then define the \emph{cost of robustness} as:
\[
\mu = \overline{r}-\overline{c}.
\]

It follows directly that $\mu \ge \delta$. Furthermore, from Theorem~\ref{thm:Nbound}, we can infer the existence of a feasible solution to $LP_1$ such that equality holds in Eq.~\eqref{eqn:lp1:cond3} (implying that $\mu = \delta$), and that $K \le \|B^{-1}\|_2 \cdot (c_1 + O(\delta))$. Therefore, when $\delta \to 0$, the mechanism achieves $\Theta\left(\frac{\delta}{c_1 \cdot \|B^{-1}\|_2}\right)$-compactness and is robust against a $\Theta\left(\frac{\delta}{c_1 \cdot \|B^{-1}\|_2}\right)$ fraction of adversarial verifiers.
Formally, we have:
\begin{theorem}
    For $\eta < \frac{1}{ g_2(k,p_1)\|B^{-1}\|_2}$, in order to ensure $\eta$ compactness, we only need a margin and cost of robustness
    \begin{equation}
        \mu =\delta \le \frac{\eta c_1g_1(k,p_1)\|B^{-1}\|_2}{1-\eta g_2(k,p_1)\|B^{-1}\|_2}.
        \label{eqn:mu}
    \end{equation}
    \label{thm:cost}
\end{theorem}
The proof is deferred to Appendix~\ref{app:thm:cost}.

\section{Byzantine Reduction for Inaccurate Beliefs and Priors}
\label{sec:inacc}

In Section~\ref{sec:robustness}, we discussed the construction of Byzantine-robust mechanisms for decentralized verification games in the presence of a small fraction of malicious players, assuming the prior knowledge of the distribution of $\theta$ is accurate. However, due to the possible existence of cheating provers, as we discussed in Section~\ref{sec:model}, there is actually a ``small but unknown'' $\eps$ probability that a proof is invalid, which is regarded as $0$ in the derivation of previous sections. While we may argue that the $\delta$ margin can indeed ensure the IA properties in the presence of ``sufficiently small'' perturbations of prior distributions, we still need explicit and quantitative results \added{of \emph{distributional robustness}} on the relations between $\delta$ and $\eps_0$ to show the practical robustness of our mechanism against dishonest provers.

In this section, we show that the $(\delta,K)$-compactness criterion is not only effective for robustness against malicious players, but also for robustness against inaccurate knowledge of prior distributions. Formally, a $(\delta,K)$-compact CTF-PP mechanism maintains its IA guarantees even if the actual distribution of $\theta$ has an $O(\delta/K)$ \emph{total variation} (TV) distance from the $P(\theta)$ we use in the construction of scoring rules, with a \emph{Byzantine reduction} argument that reduces inaccurate beliefs to the existence of malicious players. With this technique, we not only show the general robustness of our design against malicious verifiers and inaccurate priors ($\epsilon>0$, which represents the fraction of malicious provers), but also show the generality of our $(\delta,K)$-compactness criteria and Byzantine-robustness as general conceptual guidelines of robust peer prediction mechanisms.

In the rest of this section, we refer to the TV distance when we mention the distance between distributions.

\subsection{Byzantine Reduction: Inaccurate Beliefs $\le$ Malicious Players}

Before deriving the robustness results against inaccurate priors, we first discuss the robustness properties against inaccurate beliefs. 

In the notation of Section~\ref{subsec:robust:incentive}, we suppose that the system has a perception of the environmental variable as $\hat{\phi}$, while the actual environmental variable is $\phi$. Assuming that the player $i$ observes $X_i\in S\cup \{\bot\}$, her posterior belief on the distribution of $\X_{-i}$ is $\B_i(X_i,\hat{\phi})=P(\X_{-i}|X_i,\hat{\phi})$. However, as the actual environmental variable $\phi$ is (slightly) different from $\hat{\phi}$, the actual posterior distribution of $\X_{-i}$ is $\B_i(X_i,\phi)=P(\X_{-i}|X_i,\phi)$.

In the pairwise-scoring mechanism defined in Section~\ref{section:nv}, for the environmental variable $\phi$, the expected utility of reporting $Z_i$ when observing $X_i$ is
\begin{small}
\begin{align}
    r_{X_i}(Z_i) - c(X_i) = \frac{1}{n-1}\sum_{j\ne i}\left(\sum_{Z_j\in S} P(Z_j|X_i,\phi) T_{Z_iZ_j} - c(X_i)\right).
\end{align}
\end{small}

Assuming that all other players are honest, i.e., $Z_j=X_j$, from symmetry we have
\begin{small}
\begin{align}
    r_{X_i}(Z_i) - c(X_i) 
    &= \frac{1}{n-1}\sum_{j\ne i}\left(\sum_{X_j\in S} P(X_j|X_i,\phi) T_{Z_iX_j} - c(X_i)\right)\\
    &=P(X_j|X_i,\phi) T_{Z_iX_j} - c(X_i),\label{eqn:tmptmp}
\end{align}
\end{small}
in which the $j$ in \eqref{eqn:tmptmp} can be an arbitrary player different from $i$. Intuitively, if $P(X_j|X_i,\hat{\phi})$ is close to $P(X_j|X_i,\phi)$, then even if the scoring matrix $\{T_{Z_iZ_j}\}$ is designed for the $\delta$-IA property according to $\hat{\phi}$, i.e.,
\begin{small}
\begin{equation}\label{eqn:inacc:1}
\sum_{X_j\in S}P(X_j|X_i,\hat{\phi}) T_{Z_iX_j} - c(X_i)
        \begin{cases}
        \ge \delta, &Z_i=X_i; \\
        \le -\delta, &Z_i\neq X_i.
        \end{cases}
\end{equation}
\end{small}
the margin of $\delta$ can still make the mechanism $0$-IA for the actual environment of $\phi$, i.e.,
\begin{small}
\begin{equation}\label{eqn:inacc:2}
\sum_{X_j\in S}P(X_j|X_i,\phi) T_{Z_iX_j} - c(X_i)
        \begin{cases}
        \ge 0, &Z_i=X_i; \\
        \le 0, &Z_i\neq X_i.
        \end{cases}
\end{equation}
\end{small}
Actually, if the scoring matrix $T$ is $(\delta,K)$-compact, we only need that the total variation (TV) distance between $P(X_j|X_i,\hat{\phi})$ and $P(X_j|X_i,\phi)$ is bounded below $\Theta(\delta/K)$. Formally, we have:
\begin{lemma}\label{lem:tv}
    If Eq.~\eqref{eqn:inacc:1} holds, $\max\{|T_{Z_iZ_j}|\}\le K$, and $TV_{X_j}(P(X_j|X_i,\hat{\phi}),P(X_j|X_i,\phi)) \le \frac{\delta}{2K}$, then Eq.~ \eqref{eqn:inacc:2} holds.
\end{lemma}

The proof of Lemma~\ref{lem:tv} is deferred to Appendix~\ref{app:lem:tv}. From Lemma~\ref{lem:tv} we immediately deduce the following theorem:

\begin{theorem}\label{thm:tv}
Assume that there are no rogue players in environments $\phi$ and $\hat{\phi}$. If a pairwise-scoring CTF-PP mechanism is $(\delta,K)$-compact for environment $\hat{\phi}$, and for a player $j\ne i$  it holds that the total variation distance $TV(P(X_j|X_i,\hat{\phi}),P(X_j|X_i,\phi)) \le \frac{\delta}{2K}$ for every possible observation $X_i\in S\cup\{\bot\}$, then the mechanism is $0$-IA for environment $\phi$.
\end{theorem}

Actually, we can see that if $TV_{X_j}(P(X_j|X_i,\hat{\phi}),P(X_j|X_i,\phi)) = \alpha\cdot \frac{\delta}{2K}$ for $\alpha\in[0,1]$, then the mechanism is $(1-\alpha)\delta$-IA for environment $\phi$, and hence according to Lemma~\ref{lem:byzantine}, it is still Byzantine-robust to $(1-\alpha)\cdot\frac{\delta}{2K}(n-1)$ malicious players, showing that the compactness of $\delta/K$ serves as a ``reservoir'' of robustness against both inaccurate beliefs and malicious players. Intuitively, it indicates that the robustness against a $\Delta$ fraction of malicious players \emph{implies} the robustness against a $\Delta$ level of noise in posterior beliefs. This intuition can be interpreted via \emph{coupling argument}: if the posterior distribution is different from the belief, it can be equivalently regarded as the players ``within the total variation'' reporting dishonestly. The detailed discussion is deferred to Appendix~\ref{app:coupling}. 

\subsection{Robustness for Inaccurate Priors}
\label{subsec:inacc:prior}

In this part, we look into the effects of inaccurate prior distributions on the posterior beliefs, and derive the robustness of our mechanism in the presence of inaccurate priors. Under mild assumptions, we can show that an $O(\Delta)$ TV distance between two different priors is generally equivalent to an $O(\Delta)$ TV distance between corresponding posterior beliefs; hence, a $(\delta,K)$-compact mechanism is also robust in the presence of a $\Theta(\frac{\delta}{K})$ noise in prior distributions. Formally, 
\begin{theorem}\label{thm:inacc:prior}
    Assume that environments $\phi$ and $\hat{\phi}$ have no rogue players and are identical except for different prior distributions $P(\theta|\phi)\ne P(\theta|\hat{\phi})$, and $P(X_i=1|\theta\ne 1)=0$.  If a pairwise-scoring CTF-PP mechanism is $(\delta,K)$-compact for 
    environment $\hat{\phi}$ and 
    \begin{small}
    \begin{equation}
    TV_{\theta}(P(\theta|\phi), P(\theta|\hat{\phi})) \le \frac{\delta}{4K}\cdot \min_{X_i\in S^*, \varphi \in \{\phi,\hat{\phi}\}} \left\{P(X_i|\varphi)\right\}.
\end{equation}\label{eqn:theta:TV}
    \end{small}
    then the mechanism is $0$-IA for environment $\phi$.
\end{theorem}

Here, the ``min'' represents the minimum probability that any non-dishonest observation (i.e., ``Honest'' or any flag) is observed, which is a positive constant dependent on the verification protocol. The proof is deferred to Appendix~\ref{app:proof:prior}. From this result, we particularly show that the incentive guarantees robustly hold when at most an $\eps_0=\Theta(\frac{\delta}{K})$ fraction of proofs are dishonest.

\section{Experimental Evaluation}
\label{sec:numerical}

In this section, we perform numerical experiments and compare our mechanism with existing peer prediction mechanisms to show  the effectiveness of our design.

\subsection{Benchmarks} 
\label{subsec:exp:benchmark}

\textbf{Elicitation environments.} To evaluate the performance of our mechanism compared to existing mechanisms, we introduce two information elicitation environments. 

The first environment is ``Coin'' in which an unfair coin may have a type $\theta=h$ with head probability $P(X_i=H|\theta=h)=0.8$ or type $\theta=l$ with head probability $P(X_i=H|\theta=l)=0.2$, the principal prior is $P(\theta=h)=0.4, P(\theta=l)=0.6$, and the observation cost is $c_H=c_T=1$. This environment represents a standard scenario of information elicitation.

The second environment is Proof-of-Learning (PoL) as described in Appendix~\ref{sec:exp} with principal prior $P(\theta=0)=\frac{1}{2}, P(\theta=F_1)=P(\theta=F_2)=\frac{1}{4}, P(\theta=1)=0$, observation matrix as Table~\ref{table:observe} and observation costs $c(0)=\frac{1}{3},c(F_1)=c(F_2)=c(1)=2$. Here, ``$0$'' means ``valid'', ``1'' means ``invalid'' and $F_1,F_2$ stand for flags.

\textbf{Player strategies.} In our experiments, we consider three player strategies: ``Honest'', ``Lazy'', and ``Adversarial''. 

In the Honest strategy, the player honestly observes and reports her observation. 
In the Lazy strategy, the player does not observe and reports a type that maximizes her expected reward among uninformed strategies. 
In the Adversarial (permutation) strategy, the player observes but reports a flipped type, that is: in Coin, reporting $H$ when observing $T$ and vice versa; in PoL, reporting $0$ when observing $1$, reporting $F_1$ when observing $F_2$, and vice versa.

\textbf{Experiment schemes.} We perform two experiments to evaluate the basic performance and robustness to inaccurate priors.
In the first experiment, we simulate a $2$-verifier DVG in which the principal prior is accurate,
 and the peer is Honest.
In the second experiment, we simulate a $2$-verifier DVG in which the prior has an $\eps$ distance from the principal and the peer is Honest.

\textbf{Evaluation rubrics.} In each experiment, we evaluate three properties of our mechanism compared to baseline mechanisms: incentive guarantees, variance, and budget, described as follows:

\begin{itemize}
\item Incentive guarantees: We want to ensure that the Honest strategy yields non-negative utility, while the Lazy strategy (and Adversarial, if possible) yields non-positive utility.
\item Variance: As real-world players are typically risk-aversive, we report the standard deviation of players' net utilities given they play Honestly. (This concept is also studied by \citet{10.1145/3589334.3645679}.)
\item Budget: We report the expected amount of money the system needs to pay players. To save the cost, it is preferably as little over expected verification cost as possible.
\end{itemize}

\subsection{Baselines} 

To evaluate the performance of our design, we compare its incentive guarantees, variance, and budget to the baseline mechanisms as follows. In the third experiment, we compute the scores via pairwise average as described in Section~\ref{section:nv}.

\begin{itemize}
    \item Simple Agreement (SA): The player is rewarded $+r$ if her report agrees with the peer, and $-r$ otherwise.
    \item Logarithmic Scoring Rule (Log): The player is rewarded $\log P(X_{-i}=Z_{-i}|X_i=Z_i)$ when she reports $Z_i$ and her peer reports $Z_{-i}$.
    \item Pointwise Mutual Information Scoring Rule (PMI): The player is rewarded $\log \frac{P(X_{-i}=Z_{-i}, X_i=Z_i)}{P(X_{-i}=Z_{-i})\cdot P(X_{i}=Z_{i})}$ when she reports $Z_i$ and her peer reports $Z_{-i}$ \cite{zheng2024truthful}. 
    \item DMI Mechanism (DMI): The multi-task mechanism proposed by \citet{kong2023dominantly}. As the DMI mechanism needs at least $2k$ tasks in which $k$ is the number of different observations, we perform the experiments with $2k$ and $10k$ tasks to show its performance with different number of tasks. 
\end{itemize}

To ensure a fair comparison, we apply an affine transformation $f(x)=ax+b$ to the scores of each baseline mechanism. We choose the smallest possible $a$ (and corresponding $b$) such that the incentive-alignment guarantees hold (for DMI, we do not enforce UniIC and allow adversarial utilities to be positive), thereby giving each baseline the best opportunity to minimize variance and budget.\footnote{ In the multi-task DMI mechanism, the ``budget'' we report is the budget per task and the ``variance'' is divided by $\sqrt{T}$ in which $T$ is the number of tasks. If we run a single-task mechanism $T$ times, the standard deviation of total utility is $\sqrt{T}$ times the standard deviation for a single task. Hence, we divide the standard deviation by $\sqrt{T}$ for fair comparison.} Furthermore, in the experiment for inaccurate priors, \emph{{we assume that the accurate prior is known by the system for all baselines (but not in our design)}}, and is the same across all tasks for the DMI baseline. While these assumptions may not be realistic, we are allowing these baselines to operate under their most favorable conditions.

\subsection{Experiment Results}
\label{subsec:exp:res}

In this section, we show the results of the first experiment (accurate prior, honest peer) in Tables~\ref{table:exp:coin}-\ref{table:exp:pol}, and defer the second to Appendix~\ref{app:exp2}.

In the Coin benchmark, we show that if we do not enforce a $\delta$ margin, our mechanism achieves optimal budget that equals the observation cost, and also achieves the smallest variance among all listed mechanisms, showing that our objective of magnitude minimization also \emph{implicitly minimizes the variance} as it is upper bounded by the magnitude of scores. On the other hand, the PMI mechanism is the most competitive among all the baselines. Meanwhile, the DMI mechanism, though achieving desirable budget and prior-free incentive compatibility, has the worst variance and significantly worse compactness (robustness) than our mechanism under the same $\delta$ and budget. Furthermore, even though we enforce a $\delta$ margin for honest and lazy strategies, its inherent non-permutation-proof property renders it subject to adversarial reports. Hence, our mechanism achieves better performance than the DMI mechanism in the standard setting.

The results for the PoL benchmarks are similar, in which the PMI mechanism is also the most competitive among all the baselines. Nevertheless, the DMI mechanism does not work in the case of $\eps=0$ (no cheating provers) as the reward is always zero. Particularly, the DMI mechanism requires a full-rank ``answer matrix'' to distribute non-zero rewards, so the verifiers' rewards would be zero unless at least one of the tasks are done by a cheating prover, whether or not ``flags'' are adopted. Hence, it \emph{only rewards the verifiers when cheats are detected, similar to opML \cite{conway2024opml}}, which is not robust for $\eps \to 0$ and fails to resolve the Verifier's Dilemma.

\begin{table*}

\centering
\begin{small}
\begin{tabular}{|c|c|c|c|c|c|c|}
\hline
& Budget     & Variance & Compactness & Honest Utility & Lazy Utility & Adversarial Utility  \\ \hline
 Ours $(\delta=0)$ &  $\mathbf{1.00}$  & $\mathbf{2.87}$ & $0.000$ &  $0.00$ & $0.00$ & $-2.13$  \\ \hline
Ours $(\delta=0.2)$ &  $1.20$  & $3.92$ & $\mathbf{0.038}$ &  $0.20$ & $-0.20$ & $-2.69$  \\ \hline
  SA &  $1.57$  & $6.11$ & $0.000$ &  $0.57$ & $0.00$ & $-4.18$  \\ \hline
   Log &  $1.38$  & $4.26$ & $0.000$ &  $0.38$ & $0.00$ & $-2.65$  \\ \hline
   PMI &  ${1.12}$  & ${3.29}$ &  $0.000$ & $0.12$ & $0.00$ &  $-2.37$ \\ \hline
DMI ($2k, \delta=0$) &  $\mathbf{1.00}$  & $18.47$ & $0.000$ & $0.00$ & $0.00$ & $0.00$ \\ \hline
DMI ($2k, \delta=0.2$) &  $1.20$  & $25.86$ & $0.001$ & $0.20$ & $-0.20$ & \textcolor{red}{$\mathbf{0.20}$} \\ \hline
DMI ($10k, \delta=0$) &  $\mathbf{1.00}$  & $6.78$ & $0.000$ & $0.00$ & $0.00$ & $0.00$ \\ \hline
\end{tabular}
\end{small}
\caption{Experiment Results for Coin Benchmark}\label{table:exp:coin}
\end{table*}

\begin{table*}

\centering
\begin{small}
\begin{tabular}{|c|c|c|c|c|c|c|}
\hline
& Budget     & Variance & Compactness &  Honest Utility & Lazy Utility & Adversarial Utility  \\ \hline
 Ours $(\delta=0)$ &  $\mathbf{0.75}$  & $\mathbf{2.56}$ & $0.000$ & $0.00$ & $0.00$ & $-2.85$  \\ \hline
Ours $(\delta=0.2)$ &  ${0.95}$  & ${3.65}$ & $\mathbf{0.027}$ & $0.20$ & $-0.20$ & $-3.30$  \\ \hline
  SA  &  \multicolumn{6}{|c|}{\textcolor{red}{\textbf{{(Infeasible)}}}} \\ \hline
 Log &  $4.94$  & $23.06$ & $0.000$ & $4.19$ & $0.00$ & $-\infty$  \\ \hline
 PMI &  $1.25$  &  $3.84$ &  $0.000$ & $0.50$ & $0.00$ &  $-\infty$ \\ \hline
DMI &   \multicolumn{6}{|c|}{\textcolor{red}{\textbf{{(Infeasible)}}}} \\ \hline
\end{tabular}
\end{small}
\caption{Experiment Results for PoL Benchmark}\label{table:exp:pol}
\end{table*}

\section{Discussion}

In this paper, we develop a theoretical framework for the decentralized verification game on decentralized validation protocols and get theoretical results to robustly resolve the Verifier's Dilemma in a fully decentralized environment, \added{potentially reinforcing the backbone of decentralized AI incentive systems}. On the other hand, we also explore the design of peer prediction mechanisms with broader agent strategy spaces and more general settings 
and dive into its robustness issue. In future work, we will improve and broaden the study in the following aspects:

\begin{enumerate}
\item Although the PoW/PoS protocols minimize the influence of Sybil attacks, they do not eliminate them completely. In our future work, we will look into more precise economic models w.r.t. PoW/PoS protocols and discuss the resilience against Sybil attacks of our mechanisms.
\item While this paper is mainly on the elicitation of truthful verification results, we will also develop back-end voting/aggregation mechanisms that (optimally) make decisions on whether to accept the proof.
\item Beside the applications of blockchain and decentralized verification games, we will explore broader scopes of potential applications of Byzantine-robust peer prediction for decentralized consensus (e.g. Ethereum slashing), and human feedback elicitation for RLHF and AI model training/inference.
\end{enumerate}

\bibliographystyle{apalike}
\bibliography{references}

\ECSwitch
\begin{center}
    \Large \textbf{Appendix}
\end{center}

\section{Introduction of Decentralized AI Verification Protocols}
\label{app:DeAI}

Amid the rapid development of AI technologies in the LLM era, \emph{decentralized AI (DeAI)} has emerged as a promising paradigm that aims to deploy AI infrastructure on decentralized platforms such as blockchains \cite{wang2024sok}. A key motivation behind DeAI is to ensure the \emph{trustworthiness} of AI systems—specifically, to verify that training and inference processes are faithfully executed and free from adversarial tampering.

In addition to mitigating the risk of malicious attacks on AI models \cite{BytedanceAttack}, DeAI also addresses growing concerns about \emph{AI safety} \cite{bengio2024managing}. While much of the existing AI safety literature focuses on \emph{internal} risks—particularly issues of \emph{alignment} \cite{qi2024safety,chua2024ai}—these approaches typically assume that the models are correctly trained and executed. However, due to the black-box nature of AI models, \emph{external} risks arise—namely, that model developers may have incentives to manipulate the system for their own benefit, especially when model outputs influence high-stakes decisions. Thus, verifying and certifying the \emph{integrity} of AI models—that they are properly trained and function as intended—is essential.

Centralized AI corporations may be incentivized to manipulate AI systems in the absence of transparency. In contrast, decentralized verification offers a trustless approach to ensure model integrity. Therefore, DeAI plays a crucial role in mitigating external risks by ensuring model integrity through decentralized technologies \cite{saleh2024blockchain}.

From a methodological perspective, existing approaches to decentralized verification of AI models can be broadly categorized into two types: \emph{cryptographic} and \emph{game-theoretic} methods. Cryptographic methods aim to provide strong, provable guarantees of training and inference integrity, typically through mechanisms such as zero-knowledge proofs (e.g., zkML, \citet{chen2024zkml}) to ensure verifiability without revealing sensitive information. However, these methods often incur substantial computational overhead (typically exceeding 1000x) which severely undermines system efficiency and poses a significant barrier to practical deployment, particularly for large-scale models.

Alternatively, game-theoretic approaches aim to leverage economic incentives to ensure that all participants (e.g., trainers and verifiers) act honestly as a strategic equilibrium behavior—namely, that truthful actions constitute a Nash equilibrium. 

\subsection{opML: Optimistic Machine Learning for Model Inference}

A representative example is the mechanism of \emph{opML} (Optimistic Machine Learning, \citet{conway2024opml}), which secures the correctness of AI model \emph{inference} via the \emph{Optimistic Rollup} framework \cite{armstrong2021ethereum}. In this context, the term ``optimistic'' refers to the principle that \emph{all submitted computations are presumed valid unless proven otherwise.} Verifiers are thus incentivized to verify the outputs and are rewarded for successfully identifying incorrect computations.

Specifically, when a verifier verifies a submitted ML task, they re-execute the computation and compare the results:
\begin{itemize}
    \item If the results match, the task is accepted as valid.
    \item If the results do not match, a \emph{committee voting} procedure is invoked, wherein a designated committee determines the validity of the task through majority vote.
\end{itemize}

\textbf{Economic issues.} Due to the optimistic assumption of correctness, it is essential that verifiers in opML are sufficiently incentivized to invest the necessary computational resources for verification. Hence, the mechanism should reward the efforts verifiers make to offset the computational cost.

Assuming that the opML mechanism works as expected, we can expect that most provers act honestly and an overwhelming majority of submitted computations are valid. Then, we may expect a \emph{lazy} verifier to accept every proof without actual verification, unless the reward for detecting an invalid proof makes a difference. 

To simplify the discussion, we assume that the committee voting can always correctly determine if the proof is valid. From the perspective of the verifier,
\begin{itemize}
    \item If she acts honestly, she accepts a valid proof with $(1-\eps)$ probability and detects an invalid proof with $\eps$ probability.
    \item If she acts lazily, she accepts a valid proof with $(1-\eps)$ probability and accepts an invalid proof with $\eps$ probability.
\end{itemize}

We see that the outcomes only differs in the scenario that the prover cheats, which comes with a small probability of $\eps$. Hence, if the verification cost is $C$, the reward $R$ of detecting and penalty $L$ for failing to detect must sum up to $R+L\ge \frac{C}{\eps}$ to incentivize honest verification. Actually, \citet{conway2024opml} show that for given $\{R,L,C\}$, the protocol would suffer an $\eps=\frac{C}{R+L}$ rate of invalid computation at Nash equilibrium, which resembles the Verifier's Dilemma and undermines the trustworthiness of the ecosystem particularly when the verification cost $C$ is substantial.

\textbf{Attention challenges.} To address the Verifier's Dilemma, \citet{conway2024opml} propose a mechanism known as \emph{attention challenges}, which operates as follows. Suppose the prover has address $A_s$ and the output is $f(x)$:

\begin{itemize}
    \item The prover first reveals the hash value $H(f(x), A_s)$ and issues a challenge to all verifiers $v$ such that $H(f(x), A_v) < T$, where $T$ is a predefined threshold.
    \item After a fixed time window, the prover reveals the full output $f(x)$ and computes $H(f(x), A_v)$ for each verifier. Any verifier for whom $H(f(x), A_v) < T$ but who did not respond is marked as non-participating.
    \item If the submitted output $f(x)$ is ultimately deemed valid, all such non-participating verifiers are penalized, and a portion of their penalties is awarded to the prover.
\end{itemize}

Nevertheless, this design is based on the assumption that computing (and verifying) $H(f(x), A_v)$ is computationally negligible. While this assumption holds for model \emph{inference} tasks---where $f(x)$ is a simple prediction vector or classification result, it fails for training tasks, where $f(x)$ represents an entire trained model and can be prohibitively large. In this case, the Verifier's Dilemma occurs to the prover in turn.

\subsection{Proof-of-Learning (PoL): Lightweight Verification For Model Training} 

Whereas the designs of zkML and opML mainly apply to ML model inference, additional challenges occur in the development of verification protocols for ML training. Besides the fact that the training process is substantially more computationally intensive than inference, the output of the training task---the trained models---also have large sizes and even simple operations on them (e.g., hashing or comparison) take non-negligible computational costs, so that it would be harder to obtain ``cheaply-shared ground truths'' (like the $H(f(x),A_v)$ discussed above) to bypass the Verifier's Dilemma via cheap verification.
Furthermore, the re-execution method in opML would incur an at least 1x computational overhead. For ML training tasks with heavy computational costs, we are still motivated to lower this computational overhead. 

In light of this, \citet{jia2021proof} propose a ``vanilla'' Proof-of-Learning (PoL) mechanism in which the prover is supposed to train the model while leaving checkpoints during the training process, and the verifier chooses the ``most suspicious'' parts of the training process to verify via re-execution. Nevertheless, the vanilla PoL leaves a substantial gap to decentralized AI verification as it assumes the credibility of verifiers (which is unrealistic especially in the presence of the Verifier's Dilemma), and its criteria of ``most suspicious'' parts is also subject to adversarial attacks \cite{zhang2022adversarial,fang2023proof}, resembling the Goodhart's Law (\emph{``When a measure becomes a target, it ceases to be a good measure''}, \citet{goodhart1984problems}). 

To adapt PoL for decentralized AI applications, \citet{zhao2024proof} introduce a refined mechanism called \emph{incentive-secure PoL}, which replaces selective re-execution with \emph{random sampling}. They demonstrate that, under mild assumptions, this protocol satisfies \emph{incentive-security} for rational provers: dishonest behavior is detected with high probability unless the prover deviates during only a negligible fraction of training steps—insufficient to meaningfully affect performance or yield economic gain. This approach retains the lightweight nature of PoL while aligning with the incentive constraints of decentralized verification.

In addition, \citet{zhao2024proof} propose a \emph{capture-the-flag} mechanism to further strengthen verifier engagement. Here, flags—introduced as randomness by the prover—serve as verifiable tokens that honest verifiers are incentivized to detect and report, even when all proofs are valid. This incentivizes verifier efforts regardless of adversarial behavior.

The incentive-secure PoL protocol operates as follows:

\begin{itemize}
    \item The prover runs a multi-stage stochastic training process (e.g., stochastic gradient descent), recording model weights and commit hashes after each stage. At a subset of stages, cryptographic \emph{flags} are randomly inserted and committed using distinct random seeds.
    \item The verifiers randomly choose a small fraction of stages and request the prover to reveal the model weights before and after each selected stage.
    \item The prover responds by revealing the requested model weights.
    \item Verifiers check the correctness of these stages and privately commit to two reports: (1) whether the proof is accepted and (2) which flags, if any, they detect.
    \item The prover reveals the list of inserted flags.
    \item Verifiers reveal their reports and detected flags.
    \item Provers and verifiers receive rewards or penalties based on the consistency of reports and flag detections.
\end{itemize}

While \citet{zhao2024proof} do not conduct a formal game-theoretic analysis of the verifiers’ scoring rules, our study fills this theoretical gap—especially in the context where invalid proofs can only be detected probabilistically. We provide a rigorous incentive analysis that characterizes equilibrium behavior under this capture-the-flag framework, offering a foundation for incentive-aligned training verification in decentralized AI.

\section{Discussion on Strong-SCP Peer Prediction Mechanisms}
\label{app:strong:scp}

In the scope of peer prediction mechanisms, it is assumed that players report in a way that maximizes their expected utilities w.r.t. their \emph{beliefs} on other players' reports. Hence, different beliefs may lead to different estimations of utilities and different strategies. In the canonical setting of individual non-colluding players, their beliefs are the Bayesian posteriors conditioned on their observations, i.e. $P(\X_{-i}|X_i)$.

When the collusions occur, nevertheless, the beliefs may differ in different settings as the level of information and utility sharing may vary. In the notion of side-contract-proofness (SCP) in the transaction fee mechanism design \citep{shi}, it is assumed that the utilities are transferable via side payments, so that rather than a local Pareto improvement that weakly benefits everyone, a successful collusion only needs to improve the total utility of all players in the colluding party. 

While we inherit the SCP notion that allows side payments, in the context of information elicitation, there can also be different settings on whether the players' beliefs are shared or not. In the weak-SCP notion, we consider the non-sharing-belief setting that the players estimate their utilities based on their individual observations $P(X_i|\X_{-i})$, and we show in Theorem~\ref{thm:scp} that weak-SCP can be achieved via our design that optimizes $(\delta,K)$-compactness. In contrast, a ``strong-SCP'' notion considers the sharing-belief setting with players estimating their utilities based on the collective observations of the colluding party. Namely, their beliefs on the reports inside the party are just their true reports, and their beliefs on the reports outside the party are computed by $P(\X_{-\calC}|\X_{\calC})$. 

Nevertheless, a series of difficulties occur in the design for strong-SCP mechanisms.

\if
To organize the relations among different collusion-proofness notions, an interesting question is whether strong-SCP implies weak-SCP. While it is less obvious than ``strong-SCP $\implies$ quasi-strong-SCP'', the answer is true. Formally, we have:
\begin{theorem}
    A strong-SCP mechanism for decentralized verification games is weak-SCP.
\end{theorem}
\begin{proof}
Assume that a mechanism is not weak-SCP, then there exists a case in which 

\end{proof}
\fi

\subsection{Challenges in the Design of Strong-SCP Mechanisms}

\textbf{Free-riding at low noises.} Consider the scenario when the observation noise is low, i.e., the players observe the true type $\theta$ with high probability. Then in a colluding party, one observation is sufficient to secure a high confidence that other members would also observe that result, which is likely to be the ground-truth. Hence, it is rational for other players to lazily report that result too, saving the observation cost (which is typically high in ML verification contexts).

\textbf{Preference towards agreeing reports.} Even if the observation cost is low enough to keep the players willing to do active observation, as a wide scope of practically used peer prediction mechanisms, e.g., the correlated agreement (CA) mechanism \citep{shnayder2016informed} and mutual-information-based mechanisms \citep{chen2020truthful}, typically rewards agreeing reports, there can be a tendency that all colluding players report the same even if they observe differently, violating strong-SCP properties. This challenge generalizes to the family of \emph{pairwise-scoring} mechanisms, in which the colluding party's total utility is approximately linear to their average report when $n$ is large.\footnote{The approximate linearity property is actually not restricted to pairwise-scoring mechanisms, so the results can potentially be further generalized in practice.} Hence, it is impossible to design a pairwise-scoring peer prediction mechanism that satisfies a ``strict'' strong-SCP property with sensitivity guarantees (truthful reporting yields at least $h$ more utility than reporting the same when disagreement occurs). Formally, we have:

\begin{theorem}
\label{thm:app:sscp}
    In an $n$-player decentralized verification game (DVG), for any pairwise-scoring mechanism with a pairwise scoring matrix $T$ such that the reward of player $i$ is given by
    \begin{equation}
        R_i(Z_i,\Z_{-i}) = Z_i' T \overline{\Z_{-i}},
        \label{eqn:scp:scoring}
    \end{equation}
    and $-K\le T \le K$, it holds that: 
    
    \noindent \textbf{(i)} For a collusion party $\calC=\{i_1,\cdots,i_c\}$ with observations $\X_\calC = \{X_{i_1},\cdots,X_{i_c}\}$, there exists an all-same report $\Z_{\calC}^* = \{X_{i^*},\cdots,X_{i^*}\}$ such that $i^* \in \calC$ and
    \begin{align}\label{eqn:scp:impossibility}
        &\E_{\Z_\calC=\Z^*_\calC,~\Z_{-\calC}\sim P(\X_{-\calC}|\X_{\calC})}\left[\sum_{i\in \calC}R_i(X_i^*,\Z_{-i})\right] \nonumber\\
        &\ge \E_{\Z_\calC=\X_\calC,~\Z_{-\calC}\sim P(\X_{-\calC}|\X_{\calC})}\left[\sum_{i\in \calC}R_i(X_i,\Z_{-i})\right] - h,
    \end{align}
    in which
    \begin{equation}
        h= \frac{2c(c-1)}{n-1}K.
    \end{equation}
    \noindent \textbf{(ii)} If for any types $s_1,s_2\in S$, $s_1\ne s_2$, it holds that
    \[
        T_{s_1s_1} \ge T_{s_1s_2},
    \]
    i.e., the scoring rule favors agreement, then Eq.~\eqref{eqn:scp:impossibility} holds for $h\le 0$. 
    
    Furthermore, if it is possible for the colluding party $\calC$ to observe $\X_\calC$ with two different observations $X_i,X_j$ s.t. $T_{X_iX_i}>T_{X_iX_j}$, then Eq.~\eqref{eqn:scp:impossibility} holds for $h<0$, indicating that the mechanism is not strong-SCP.
\end{theorem}

The proof is deferred to Appendix~\ref{app:sscp}. We can interpret part (i) as that: when $c\ll n$, the approximate linearity implies the impossibility to disincentivize reporting the same type with significant incentive margins, as assuming the expected reward of an honest player is $\Theta(K)$, the expected reward of the party is $\Theta(cN)\gg h$. Furthermore, part (ii) shows that as long as the mechanism favors agreement (as most existing peer prediction mechanisms typically do), this type of collusion will be strictly profitable, rendering them non-strong-SCP. While we conjecture that it may be generally impossible to design a strong-SCP single-task peer prediction mechanism under mild assumptions, Theorem~\ref{thm:app:sscp} shows that if it is actually possible, we need to bypass the approximate linearity and may need extremely tricky designs. On the other hand, other possible approaches to address the collusion issue in the shared-belief setting may include:
\begin{itemize}
    \item Multi-task settings: assigning different task sets to different players and make partial-copying strategies unprofitable.
    \item Aggregation design: while failing to prevent collusion in the front-end elicitation phase, it may still be possible to design aggregation mechanisms to minimize its impact to the back-end decision-making (e.g., whether to accept the proof/block/etc.)
\end{itemize}
Nevertheless, we can notice that for the phenomenon that colluding players tend to copy the same report, its practical effect on the functionality of back-end decision-making could be rather ``benign'' as at least one of them did the honest observation, which still bypasses the Verifier's Dilemma and can be informationally sufficient particularly in the low-noise scenario: in the back-end perspective, we can also regard such colluders as one player with more voting power in an almost unanimous voting. We leave detailed analyses and discussions on these aspects for future work.

\section{A Coupling Interpretation of Byzantine Reduction}
\label{app:coupling}

\subsection{Coupling Argument and Total Variation Distance}

In probability theory and statistics, \emph{coupling} is a technique to compare characteristics of two distributions, especially when they are close to each other. In general, for two given distributions $D_1,D_2$, we can construct two dependent random variables $V_1,V_2$ a with a joint distribution $P(V_1,V_2)$ such that the marginal distributions satisfy
\[
    P(V_1)=D_1, P(V_2)=D_2.
\]

While the construction of the joint distribution $P(V_1,V_2)$ is not unique, in the particular case that $D_1$ and $D_2$ are close to each other, we would like to make $V_1=V_2$ with a high probability. 

As an intuitive interpretation, we can regard that in the ``main world'' $\Omega_1$, a random event $V_1$ happens according to $D_1$; in a parallel ``alternative world'' $\Omega_2$, the event is tampered to $V_2$ which has a different distribution $D_2$. We can imagine that such magical manipulation is costly, so that the manipulator would like to tamper as little as possible, i.e., minimize $P(V_1\ne V_2)$ as long as $V_2\sim D_2$. Actually, this model falls into the scope of \emph{optimal transport}. From optimal transport theories, it holds that
\begin{equation}
    \min_{V_1\sim D_1, V_2\sim D_2} P(V_1\ne V_2) = TV(D_1,D_2).
    \label{eqn:tv}
\end{equation}

\textbf{Example.} An academic institute has recently recruited $50$ new tenure-track assistant professors, among which $5$ are expected to get tenure, yielding a tenure rate of $10\%$. However, due to a sudden cut in funding, the tenure rate has to be lowered to $6\%$, hence some of the would-be decisions have to be changed. From Eq.~\eqref{eqn:tv}, the minimum number of changed decisions is
\begin{equation}
    50 \cdot TV(Bern(0.10),Bern(0.06)) = 50\cdot 0.04 = 2.
\end{equation}

In fact, to change the fewest decisions, the tenure decisions for $2$ unfortunate candidates will be revoked.

\subsection{Interpretation for Robust Peer Prediction}

In the context of (Byzantine)-robust peer prediction for $n$ players, we consider a mechanism that is $(\delta,K)$-compact for the environment $\hat{\phi}$ that has no rogue players. From Theorem~\ref{thm:scp} we know that this mechanism keeps the $0$-IA incentive guarantee even if an arbitrary subset of at most $\frac{\delta}{2K}(n-1)$ players, i.e. a $\frac{\delta}{2K}$ fraction of other players, become malicious. For simplicity of discussion, we assume $n\to\infty$.

We assume that in the main world, the environment is $\hat{\phi}$. Conditioned on player $i$ observing $X_i$, the (expected) distribution\footnote{We consider the distribution ensemble-wise, i.e., among all possible ground-truth $\theta$'s. The term ``(conditional) distribution'' in this part is always interpreted in this way.} of other players' observations is $P(X_j|X_i,\hat{\phi})$. From the Byzantine robustness results, the $0$-IA guarantee holds as long as at least an $1-\frac{\delta}{2K}$ fraction of them report honestly.

Then, we consider the alternative world in which the environment is $\phi$. Assuming that the player $i$ still observes $X_i$, the distribution of other players' observations is $P(X_j|X_i,\phi)$. From the coupling argument discussed above, a minimum of $TV_{X_j}(P(X_j|X_i,\hat{\phi}),P(X_j|X_i,\phi))$ fraction of other players have different observations between two worlds.

Now we assume that all other players report honestly in the alternative world, while in the main world, they also report their observations in the alternative world. Then we see that in the main world,  an exact $TV_{X_j}(P(X_j|X_i,\hat{\phi}),P(X_j|X_i,\phi))$ fraction of other players are reporting dishonestly. Hence in the main world, the $0$-IA guarantee holds as long as $TV_{X_j}(P(X_j|X_i,\hat{\phi}),P(X_j|X_i,\phi))\le \frac{\delta}{2K}$.

On the other hand, in the alternative world all players are reporting honestly, but the actual environment is $\phi$ instead of $\hat{\phi}$. Since all other players report identically in two worlds, in the perspective of player $i$, the $0$-IA guarantee still holds in the alternative world as long as $TV_{X_j}(P(X_j|X_i,\hat{\phi}),P(X_j|X_i,\phi))\le \frac{\delta}{2K}$. 
Hence we see that the Byzantine-robustness against a $\frac{\delta}{2K}$ fraction of malicious players implies the tolerance of a $\frac{\delta}{2K}$ error of posterior beliefs. 

An intuitive interpretation is that, from the perspective of a (self-centric) player 
$i$, even if the actual posterior distribution of others’ observations differs from her belief, she can interpret this difference as arising from some players reporting dishonestly. In this interpretation, the player regards her belief (the ideal posterior distribution) as correct, while the fraction of players causing the discrepancy corresponds to the error measured by the total variation distance. This ensures that as long as the error remains below the tolerance threshold $\frac{\delta}{2K}$, the robustness guarantee holds.

\section{Demonstration of th PoL Benchmark}
\label{sec:exp}

In this section, we empirically demonstrate the process of designing a CTF-PP mechanism, for one set of parameters that is useful for practical interest.

\subsection{Construction of the Scoring Rule}

We consider the 2-verifier DVG which captures the case of one stage in \cite{zhao2024proof}. Here, we set the distribution $\theta$ as $P(\theta=F_1)=P(\theta=F_2)=\frac{1}{4}$, which means that half of all stages are flagged. Then we consider the lossy-channel model in which $\mu_1=\mu_2=\frac{1}{2}$, as each verifier independently chooses half of all stages \footnote{It is significantly more than needed, but does work.} and each flag is detected with probability $1$ when verified. According to the CTF protocol, when a cheating stage is chosen by a verifier, it has an $\frac{1}{2}$ chance to be correctly detected, and a $\frac{1}{4}$ chance to be observed as $F_1$ or $F_2$ respectively. Hence, $P(X_i|\theta)$ is shown as in Table \ref{table:observe}, and assuming $\eps=0$, we compute the marginal distribution of $X_i$ as $B_\bot = [\frac{3}{4},\frac{1}{8},\frac{1}{8},0]$.

Then, assuming $\eps=0$, from $P(X_i,X_{-i})=\sum_{\theta} P(\theta)P(X_i|\theta)P(X_{-i}|\theta)$ we can compute the joint probabilities $P(X_1,X_2)$, as shown in Table~\ref{table:joint:dist}. We accordingly compute the post-observation belief $P(X_2|X_1)=\frac{P(X_1,X_2)}{P(X_1)}$ for $X_1\ne 1$, and $P(X_2|X_1=1)=P(X_2|\theta=1)$, getting the principal belief matrix $B$ as Table~\ref{table:cond:dist2}.

\begin{table}[htb]
\centering
\begin{tabular}{|c|c|c|c|c|}
\hline
            & $X_i=0$        & $X_i=F_1$      & $X_i=F_2$ &  $X_i=1$   \\ \hline
$\theta=0$     & $1$  & $0$ & $0$ & $ 0$ \\ \hline
$\theta=F_1$   & $\frac{1}{2}$ & $\frac{1}{2}$ & $0$   &  $ 0$       \\ \hline
$\theta=F_2$ & $\frac{1}{2}$ & $0$            & $\frac{1}{2}$ &  $ 0$ \\ \hline
$\theta=1 $ & $\frac{1}{2}$ & $\frac{1}{8}$            & $\frac{1}{8}$ & $\frac{1}{4}$\\ \hline
\end{tabular}
\caption{$P(X_i|\theta)$, the observation distribution conditioned on $\theta$.}\label{table:observe}
\end{table}

\begin{table}[htb]
\centering
\begin{tabular}{|c|c|c|c|c|}
\hline
            & $X_2=0$        & $X_2=F_1$      & $X_2=F_2$ & $X_2=1$     \\ \hline
$X_1=0$     & $\frac{5}{8}$  & $\frac{1}{16}$ & $\frac{1}{16}$ & $0$ \\ \hline
$X_1=F_1$   & $\frac{1}{16}$ & $\frac{1}{16}$ & $0$ & $0$            \\ \hline
$X_1=F_2$ & $\frac{1}{16}$ & $0$            & $\frac{1}{16}$ & $0$ \\ \hline
$X_1=1$ & $0$ & $0$            & $0$ & $0$ \\ \hline
\end{tabular}
\caption{Joint probabilities  $P(X_1,X_2)$ ($\epsilon = 0$).}\label{table:joint:dist}
\end{table}

\begin{table}[htb]
\centering
\begin{tabular}{|c|c|c|c|c|}
\hline
            & $X_2=0$        & $X_2=F_1$      & $X_2=F_2$ &  $X_2=1$   \\ \hline
$X_1=0$     & $\frac{5}{6}$  & $\frac{1}{12}$ & $\frac{1}{12}$ & $ 0$ \\ \hline
$X_1=F_1$   & $\frac{1}{2}$ & $\frac{1}{2}$ & $0$   &  $ 0$       \\ \hline
$X_1=F_2$ & $\frac{1}{2}$ & $0$            & $\frac{1}{2}$ &  $ 0$ \\ \hline
$X_1=1 $ & $\frac{1}{2}$ & $\frac{1}{8}$            & $\frac{1}{8}$ & $\frac{1}{4}$\\ \hline
\end{tabular}
\caption{The principal belief matrix $B$.}\label{table:cond:dist2}
\end{table}

From the nature of the CTF mechanism, in which the observation of $F_1, F_2$ of $1$ takes twice the computational cost of a stage, we set $c(F_1)=c(F_2)=c(1)=2c$. On the other hand, the ``observation'' of a $0$ has two cases: one is that the verifier has verified the stage that is not cheated or flagged, which has a $\mu(1-\eta)=\frac{1}{4}$ probability, and one is that the verifier does not verify the stage from the random verification protocol, which has a $1-\mu=\frac{1}{2}$ probability. Hence, we have an ``amortized'' $c(0)=\frac{1}{3}c$. Without loss of generality, we set $c=1$.

With the knowledge of $B,B_\bot$ and $c$, Eqs.~\eqref{eqn:lp:1}-\eqref{eqn:lp:3} are the constraints that a desirable scoring rule for a CTF-PP mechanism should satisfy. For the robustness of our mechanism, we do not want the payments to have extremely large absolute values. Hence, we construct the linear program as:

\begin{align*}
    \quad \quad minimize~\quad~   &K &\\
    \quad \quad s.t.~\quad~        &\textup{\eqref{eqn:lp:1}-\eqref{eqn:lp:3}}, \quad -K \le T \le K.
\end{align*}

We set a margin of $\delta=0.2$, and compute a numerical solution to this LP, getting a scoring rule as shown in Table~\ref{table:scoring:rule:num}.

\begin{table}[htb]
\centering
\begin{tabular}{|c|c|c|c|c|}
\hline
            & $Z_{-i}=0$        & $Z_{-i}=F_1$      & $Z_{-i}=F_2$  &  $Z_{-i}=1$   \\ \hline
$Z_i=0$     & $+2.0690$  & $-7.1451$ & $-7.1451$ & $-2.2507$ \\ \hline
$Z_i=F_1$   & $-2.0446$ & $+6.4446$ & $-4.7421$  & $-2.0022$ \\ \hline
$Z_i=F_2$ & $-2.0446$ & $-4.7421$ & $+6.4446$  & $-2.0022$ \\ \hline
$Z_i=1$ & $-2.2000$ & $+5.8000$            & $+5.8000$ & $+7.4000$\\ \hline
\end{tabular}
\caption{$T_{Z_i Z_{-i}}=R_{i}(Z_i,Z_{-i})$ as a numerical solution.}\label{table:scoring:rule:num}
\end{table}

\subsection{Evaluation}

With this scoring rule, given verifier $-i$ acts honestly, we report the expected utility of verifier $i$ is in Table~\ref{table:utility}, assuming $\eps=0$, showing that the verifier gets a positive expected utility if and only if she verifies and reports honestly. 

\begin{figure}[htb]
  \centering
    \includegraphics[scale=0.5]{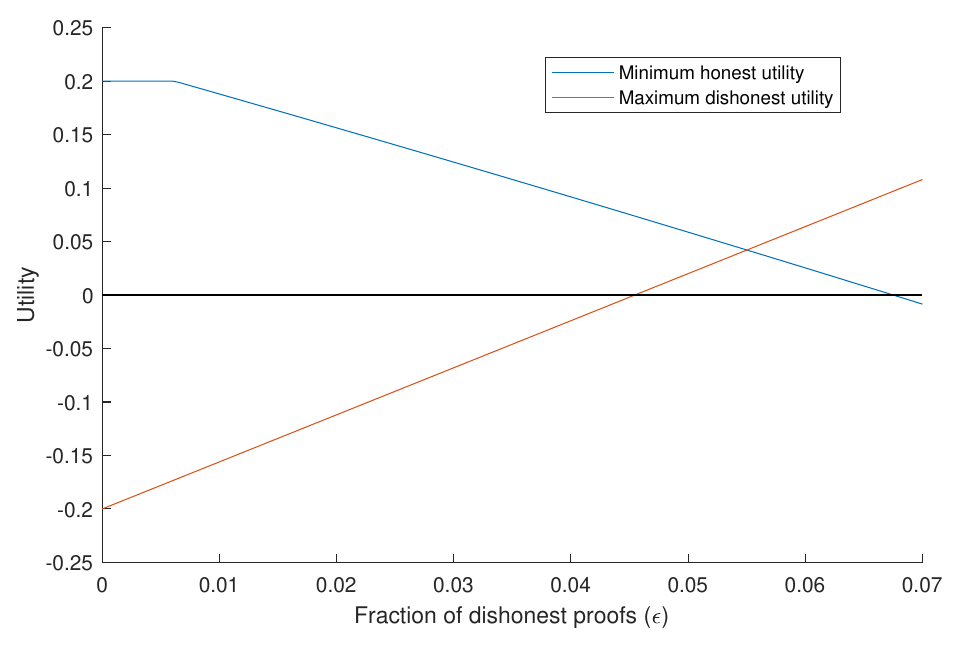}
    \caption{Verifier's expected utility, $\eps > 0$.}
    \label{fig:utility}
\end{figure}

Furthermore, we consider the case $\eps > 0$. We plot the maximum expected utility of dishonest actions and the minimum expected utility of honest actions in Figure~\ref{fig:utility}. From the plot, we show that the introduction of the margin keeps the IR, UniIC and NFL properties of our mechanism as long as $\eps < 0.045$. 
Hence, we demonstrate that our design of the CTF-PP mechanism for the 2-verifier DVG can incentivize honest verification even if there is no dishonest prover, thus bypassing the Verifier's Dilemma and achieving a pure-strategy Nash equilibrium that the prover and verifiers simultaneously act honestly.

\begin{table}[htb]
\centering
\begin{tabular}{|c|c|c|c|c|}
\hline
            & Reporting $0$        & Reporting $F_1$      & Reporting $F_2$  &  Reporting $1$   \\ \hline
Observing $0$     & $\mathbf{+0.2000}$  & $-1.8953$ & $-1.8953$ & $-1.2000$ \\ \hline
Observing $F_1$   & $-4.5381$ & $\mathbf{+0.2000}$ & $-5.3933$  & $-0.2000$ \\ \hline
Observing $F_2$ & $-4.5381$ & $-5.3933$ & $\mathbf{+0.2000}$  & $-0.2000$ \\ \hline
Observing $1$ & $-3.3144$ & $-3.3100$            & $-3.3100$ & $\mathbf{+0.2000}$\\ \hline
Uninformed & $-0.2345$ & $-1.3206$            & $-1.3206$ & $-0.2000$\\ \hline
\end{tabular}
\caption{Verifier's expected utility, $\eps = 0$.}\label{table:utility}
\end{table}

\section{Additional Experiments}
\label{app:exp2}

In this part, we report the results of the second experiment as described in Section~\ref{subsec:exp:benchmark}.

In this experiment, we consider the case of a 2-verifier DVG, in which the peer is honest but the actual prior has an $\eps$ TV distance from the principal distribution, simulating the case in which a small fraction of proofs is dishonest. 

As the PMI baseline has been shown as the most competitive among the baselines in Section~\ref{sec:numerical}, and SA and DMI are infeasible even for the noise-free case in the PoL benchmark, in the PoL benchmark we mainly compare our design with PMI. To show the potential of re-scaling PMI scoring rules for an incentive margin (and robustness), we introduce a variation of the PMI mechanism:

\begin{itemize}
\item PMI-Oracle (PMI-O): The scoring rule is computed with the actual prior (which should not have been accessible in practice) and scaled accordingly.
\end{itemize}

For convenience in the computation of utilities, we truncate the infinite entries to $\pm 20$ in the Log and PMI scoring rules. Since we have already shown in Section~\ref{sec:numerical} that the DMI mechanism always has the same adversarial utility as honest utility (which is not desired in our setting), we omit the experiments for the DMI mechanism.


Here, we let $\eps=0.01$ and $\eps=0.03$, and show the results for Coin and PoL benchmarks in Tables~\ref{table:exp:2:coin}-\ref{table:exp:2:pol2}.

From the experiment results, we can see that the introduction of $\delta$ margin can ensure positive honest utility and negative dishonest utility even in the presence of inaccurate prior of an $\eps=0.03$ TV distance. In the Coin baseline, our design with $\delta=0.2$ pays lower budget than SA and Log mechanisms, while the PMI-O baseline can achieve slightly lower budget than our design (with carefully tuned affine transformations assuming that the $\eps$ is known in advance). However, in the trickier PoL benchmark in which the Verifier's Dilemma actually occurs, our design yields better robustness and lower budgets than the PMI mechanism, even if we allow the PMI mechanism to optimally adjust the scaling factors with the accurate $\eps$. Hence, we have shown that our design achieves a more robust and cost-efficient solution for the Verifier's Dilemma than existing peer prediction mechanisms listed above.

\begin{table*}

\centering
\begin{tabular}{|c|c|c|c|c|}
\hline
& Budget   & Honest Utility & Lazy Utility & Adversarial Utility  \\ \hline
Ours ($\delta=0$) &  $1.00$  & \textcolor{red}{$-0.004$} & \textcolor{red}{$0.04$} &  $-2.13$ \\ \hline
Ours ($\delta=0.2$) &  $1.20$  & {$0.20$} & {$-0.15$} &  $-2.67$ \\ \hline
SA &  $1.57$  & $0.57$ & \textcolor{red}{$0.08$} &  $-4.18$ \\ \hline
Log &  $1.39$  & $0.39$ & \textcolor{red}{$0.07$} &  $-2.64$ \\ \hline
PMI &  $1.11$  & $0.11$ & \textcolor{red}{$0.04$} &  $-2.37$ \\ \hline
PMI-O &  $1.12$  & $0.12$ & {$0.00$} &  $-2.37$ \\ \hline

\end{tabular}
\caption{Experiment 2, Coin Benchmark, $\eps=0.01$}\label{table:exp:2:coin}
\end{table*}

\begin{table*}

\centering
\begin{tabular}{|c|c|c|c|c|}
\hline
& Budget   & Honest Utility & Lazy Utility & Adversarial Utility  \\ \hline
Ours ($\delta=0$) &  $0.99$  & \textcolor{red}{$-0.01$} & \textcolor{red}{$0.12$} &  $-2.12$ \\ \hline
Ours ($\delta=0.2$) &  $1.20$  & {$0.20$} & {$-0.04$} &  $-2.67$ \\ \hline
SA &  $1.57$  & $0.57$ & \textcolor{red}{$0.24$} &  $-4.17$ \\ \hline
Log &  $1.41$  & $0.41$ & \textcolor{red}{$0.20$} &  $-2.62$ \\ \hline
PMI &  $1.09$  & $0.09$ & \textcolor{red}{$0.12$} &  $-2.37$ \\ \hline
PMI-O &  $1.16$  & $0.16$ & {$0.00$} &  $-2.51$ \\ \hline

\end{tabular}
\caption{Experiment 2, Coin Benchmark, $\eps=0.03$}\label{table:exp:2:coin2}
\end{table*}

\begin{table*}

\centering
\begin{tabular}{|c|c|c|c|c|}
\hline
& Budget   & Honest Utility & Lazy Utility & Adversarial Utility  \\ \hline
Ours ($\delta=0$) &  $0.75$  & \textcolor{red}{$-0.004$} & $0.00$ &  $-2.85$ \\ \hline
Ours ($\delta=0.2$) &  $0.95$  & {$0.20$} & {$-0.20$} &  $-3.30$ \\ \hline
PMI &  $1.24$  & $0.49$ & \textcolor{red}{$0.01$} &  $-6.99$ \\ \hline
PMI-O &  $1.24$ & $0.49$ & $0.00$ & $-7.98$   \\ \hline

\end{tabular}
\caption{Experiment 2, PoL Benchmark, $\eps=0.01$}\label{table:exp:2:pol}
\end{table*}

\begin{table*}

\centering
\begin{tabular}{|c|c|c|c|c|}
\hline
& Budget   & Honest Utility & Lazy Utility & Adversarial Utility  \\ \hline
Ours ($\delta=0$) &  $0.74$  & \textcolor{red}{$-0.01$} & $0.00$ &  $-2.83$ \\ \hline
Ours ($\delta=0.2$) &  $0.94$  & {$0.19$} & {$-0.18$} &  $-3.28$ \\ \hline
PMI &  $1.22$  & $0.47$ & \textcolor{red}{$0.02$} &  $-6.87$ \\ \hline
PMI-O &  $1.21$  & $0.46$ & $0.00$  &  $-7.64$ \\ \hline

\end{tabular}
\caption{Experiment 2, PoL Benchmark, $\eps=0.03$}\label{table:exp:2:pol2}
\end{table*}

\section{Deferred Proofs}

\subsection{Proof of Theorem~\ref{thm:verifier:dilemma}}\label{app:veri:dil}

Assume we have such a mechanism. By the definition of Nash equilibrium, we consider a fixed verifier. Given that the prover and all other verifiers act honestly, that verifier should be incentivized to do the honest verification.

Since the prover is honest, when that verifier performs honest verification, the result should always be ``Success''. However, suppose the verifier simply reports ``Success'' without verification. In that case, the outcome is the same but the verifier saves the verification cost, so the verifier is incentivized to deviate from the honest strategy.

That leads to a contradiction, so no such mechanism exists.

\subsection{Proof of Theorem~\ref{thm:feasibility}}\label{app:feasibility}

Notice that if $B$ is invertible, for any given $W: S^2 \to \mathbb{R}$, we can compute a $T = (B^{-1}W)'$ that satisfies $W=BT'$. Since we have

\begin{align}
    B_{\bot y}&=P(X_{-i}=y)\\
    &=\sum_{x\in S}P(X_i = x)\cdot P(X_{-i}=y|X_i = x)\\
    &=\sum_{x\in S}B_{\bot x}\cdot B_{xy},
\end{align}

It holds that $B_\bot = B_\bot B$. Hence, $W_\bot = B_\bot T' = B_\bot B T'= B_\bot W$. Since $B_\bot\ge 0$, $W_\bot$ is a convex combination of rows in $W$. Therefore, we only need to construct a $W$ that satisfies Eq.~\eqref{eqn:lp:1} with non-diagonal entries small enough.

Here, for a constant $M>0$ large enough, we construct $W$ as:

\begin{align}\label{eqn:wM:1}
    W_{xx} &= c(x) + \delta,\qquad \forall x\in S; \\
    W_{xy} &= -M,\qquad \quad~~\,\forall x\in S,\quad y\in S-\{x\}.\label{eqn:wM:2}
\end{align}

Then,

\begin{align}
    W_{\bot y} &= \sum_{x\in S} B_{\bot x} W_{xy} \\
    &= B_{\bot y} W_{yy} + \sum_{x\in S-\{y\}} B_{\bot x} W_{xy} \\
    &= B_{\bot y} (c(y)+\delta) - (1-B_{\bot y}) M.
\end{align}

Since the existence of flags introduces randomness in the observation, we have $\max\{B_\bot\}<1$. Hence, denote $B^*_\bot = \max\{B_{\bot y}\}$, we only need to let

\begin{equation}
    M \ge \frac{B^*_\bot\cdot (c(y)+\delta)+\delta}{1-B^*_\bot},
\end{equation}

Then the required constraints of Eqs.~\eqref{eqn:lp:1}-\eqref{eqn:lp:3} are satisfied with a margin of $\delta$.

Then we consider the scenario that $\eps>0$ but is small enough. In this case, define $B(\eps)$ and $B_\bot(\eps)$ as the belief matrix and blind-belief matrix considering the influence of $\eps$. We can see that for any $x\ne 1$, since $\tilde{P}(X_i=x)>0$, the influences of $\eps$ on $P(X_{-i}|X_i=x)$ and $P(X_{-i})$ are upper bounded by $O(\eps)$, and because $\theta\ne 1\implies X_i\ne 1$, Eq.~\eqref{eqn:ir:2} always holds for any $\eps$. Therefore, the margin of $\delta$ ensures that the constraints are not violated as long as $\epsilon_0$ is small enough.

Finally, let $T=(B^{-1}W)'$, then $T$ is a scoring rule that satisfies the requirements.

\subsection{Proof of Proposition~\ref{prop:invert}}
\label{app:invert}

In the 2-verifier DVG, By definition $B_{xy}=P(X_{-i}=y|X_i=x)$. We sort the elements of $S$ in the order $(0,F_1,\cdots, F_{m}, 1)$. For convenience, we define the observation matrix (aka. confusion matrix) $O$ and inference matrix $E$ as:

\begin{align}
O_{xy}&=P(X_i=y|\theta=x),\\
E_{xy}&=P(\theta=y|X_i=x).
\end{align}

From the lossy-channel model, we immediately see that $O$ is an upper triangular matrix with non-zero diagonals, so $O$ is invertible.

Furthermore, given $\eps=0$, we can see that:

\begin{itemize}
    \item If $X_i=0$, then the ground truth $\theta$ may be $0$ or any flag $F_j$, and $P(\theta=0|X_i=0)>0.$
    \item If $X_i=F_j$, then $\theta=F_j$.
    \item If $X_i=1$, then $\theta=1$.
\end{itemize}
Hence, $E$ is a lower triangular matrix with non-zero diagonals, so $E$ is invertible. Then, because $X_i,X_{-i}$ are independent conditioned on $\theta$, we have
\begin{align}
    B_{X_i X_{-i}}&=P(X_{-i}|X_i)\\
    &=\sum_{\theta\in S} P(\theta|X_{i})P(X_{-i}|\theta)\\
    &=\sum_{\theta\in S}E_{X_i\theta}\cdot O_{\theta X_{-i}}.
\end{align}

Therefore,    $B=EO$.

Since $E,O$ are invertible, we deduce that $B$ is invertible.

\subsection{Proof of Theorem~\ref{thm:generalization}}
\label{app:gen:proof}

In the context of Bayesian Nash equilibrium, we can assume each verifier $j\ne i$ is honest, i.e., $\Z_{-i}=\X_{-i}$. Hence, given that verifier $i$ observes $X_i\in S \cup \{\bot\}$ and reports $Z_i\in S$, the interim expected reward is:
\begin{align}
    r_{X_i}(Z_i) &= \E\Big[Z_i'T\overline{\X_{-i}}\Big|X_i\Big] \\
    &= Z_i' T \cdot \E\Big[\overline{\X_{-i}}\Big|X_i\Big] \\
    &= Z_i' T \cdot \E\Big[\frac{1}{n-1}\sum_{j\ne i}X_j\Big|X_i\Big]\\
    &= \frac{1}{n-1}\sum_{j\ne i}Z_i'T\E\Big[X_j\Big|X_i\Big] \\
    &= \frac{1}{n-1}\sum_{j\ne i}Z_i'T\sum_{X_j\in S} P(X_j|X_i) X_j \\
    &= \frac{1}{n-1}\sum_{j\ne i}\sum_{X_j\in S} P(X_j|X_i) T_{Z_iX_j}.
\end{align}

With similar arguments as Section~\ref{section:2v}, we assume $\eps=0$, and $P(X_j|X_i)$ is the $(i,j)$-th entry of the principal belief matrix $B$ for any $j\ne i$. Hence, we have
\begin{align}
    r_{X_i}(Z_i) &= \frac{1}{n-1}\sum_{j\ne i}\sum_{X_j\in S} P(X_j|X_i) T_{Z_iX_j}\\
    &= \sum_{X_j\in S} B_{X_iX_j} T_{Z_iX_j}\\
    &= (BT')_{X_iZ_i}.
\end{align}

Hence, the linear program of Eqs.~(\ref{eqn:lp:1}-\ref{eqn:lp:3}) works equivalently for the $n$-verifier DVG when we use the linear average mechanism as Eq.~\eqref{eqn:linear:mechanism} with exactly the same incentive structure. So any incentive property satisfied in the 2-verifier mechanism $T$ is also satisfied in the linear average mechanism in Eq.~\eqref{eqn:linear:mechanism}.

\subsection{Proof of Theorem~\ref{thm:Nbound}}
\label{app:Nbound}

We first observe that any feasible solution of $LP_1(B,B_\bot,c,\delta)$ can be constructed with feasible solutions of $LP_1(B,B_\bot,c,0)$ and $LP_1(B,B_\bot,0,1)$, i.e.,

\begin{observation}
    If $(K_c,T_c)$ is a feasible solution of  $LP_1(B,B_\bot,c,0)$ and $(K_\delta,T_\delta)$ is a feasible solution of $LP_1(B,B_\bot,0,1)$, then $(K_c+\delta K_\delta,T_c+\delta T_\delta)$ is a feasible solution of $LP_1(B,B_\bot,c,\delta)$.
\end{observation}

Hence, we can estimate upper bounds of optimal $K_c$ and $K_\delta$ separately. Here we denote $\|\cdot\|_2$ as the matrix $\ell_2$-norm, and denote $W=BT'$. From the assumption in Theorem~\ref{thm:feasibility} that $B$ is invertible, $T$ can be constructed as $(B^{-1}W)'$ and it holds that $\forall x,y\in S$, $|T_{xy}|\le \|T\|_2 =\|(B^{-1}W)'\|_2\le \|B^{-1}\|_2\cdot \|W\|_2$. Hence, we can estimate the upper bounds on entrywise maximums of $T$ via $\ell_2$ norms of $W$, respectively.

For $LP_1(B,B_\bot,c,0)$, if we construct 
\begin{align}
    W_{xy} = \begin{cases}
    c,&y=x;\\
    -\frac{B_{\bot y}}{1-B_{\bot y}}\cdot c,&y\ne x.
    \end{cases}
    \label{eqn:lp1:solution}
\end{align}
Then the corresponding $T_c=(B^{-1}W)'$ obviously satisfies conditions (\ref{eqn:lp1:cond3}-\ref{eqn:lp1:cond4}). For condition~\eqref{eqn:lp1:cond5}, we have

\begin{align}
(B_\bot W)_y &= \sum_{x\in S}B_{\bot x}W_{xy} \\
&= B_{\bot y} W_{yy} + \sum_{x\in S-\{y\}}B_{\bot x}W_{xy} \\
&= B_{\bot y}\cdot 1 + (1-B_{\bot y})\cdot \bigg(-\frac{B_{\bot y}}{1-B_{\bot y}}\bigg)\\
&=0.
\end{align}

So $T_c$ is feasible for $LP_1(B,B_\bot,c,0)$, and we analyze $K_c$ later.

Before analysis for $LP_1(B,B_\bot,0,1)$, we prove a lemma:

\begin{lemma}\label{lem:bbot}
    $B_\bot'$ is an eigenvector of $B'$ with eigenvalue $1$, i.e., $B_\bot B = B_\bot$.
\end{lemma}

\begin{proof}{Proof}

    Let $j$ be an arbitrary verifier other than $j$. From the discussion of the uninformed strategy (in Section~\ref{sec:model}), we have
    \begin{align}
        B_{\bot y} &= P(X_j=y|X_i=\bot)\\
        &= P(X_j=y).
    \end{align}

    On the other hand,
    \begin{align}
    (B_\bot B)_y &= \sum_{x\in S}B_{\bot x} B_{xy}\\
    &= \sum_{x\in S} P(X_i=x)\cdot P(X_j=y|X_i=x)\\
    &= \sum_{x\in S} P(X_i=x, X_j=y)\\
    &= P(X_j=y).
    \end{align}

    Hence we have $B_{\bot y}=(B_\bot B)_y$ for $\forall y\in S$, so $B_\bot B = B_\bot$.

    \hfill $\square$
\end{proof}

From Lemma~\ref{lem:bbot}, the $LP_1(B,B_\bot,0,1)$ can be reformulated as:

\begin{align}
    &LP_2(B,B_\bot,0,1):\nonumber\\
    \quad &minimize & K \\
    &s.t. &|B^{-1}W| &\le K,\\
    &&W_{xx}&\ge 1, &\forall x\in S~ &\label{eqn:lp2:cond3}\\
    &&W_{xy}&\le -1, &\forall x\in S, &\quad y\in S-\{x\}\label{eqn:lp2:cond4}\\
    &&B_\bot W &\le -1.\label{eqn:lp2:cond5}
\end{align}

Here, we can construct
\begin{align*}
    W_{xy} = \begin{cases}
    1,&y=x;\\
    -\frac{1+B_{\bot y}}{1-B_{\bot y}},&y\ne x.
    \end{cases}
\end{align*}

From the construction we immediately see that conditions~(\ref{eqn:lp2:cond3}-\ref{eqn:lp2:cond4}) are satisfied. For condition~(\ref{eqn:lp2:cond5}), we have
\begin{align}
(B_\bot W)_y &= \sum_{x\in S}B_{\bot x}W_{xy} \\
&= B_{\bot y} W_{yy} + \sum_{x\in S-\{y\}}B_{\bot x}W_{xy} \\
&= B_{\bot y}\cdot 1 + (1-B_{\bot y})\cdot \bigg(-\frac{1+B_{\bot y}}{1-B_{\bot y}}\bigg)\\
&=-1.
\end{align}

Hence, the $W$ is feasible for $LP_2(B,B_\bot,0,1)$. Now we estimate an upper bound on $\|W\|_2$. We first show a lemma:

\begin{lemma}
    For any matrix $A$,
    \begin{equation}
        \|A\|_2 \le \sqrt{\|A\|_1\|A\|_\infty}.
    \end{equation}
\end{lemma}

\begin{proof}{Proof}
Denote $A^*$ as the conjugate transpose of $A$, which is equal to $A'$ when $A$ is real, and denote $\lambda_{max}(\cdot)$ as the maximum eigenvalue. Then it holds that
\begin{equation}
    \|A\|_2 = \sqrt{\lambda_{max}(A^*A)}.
\end{equation}

Because the maximum eigenvalue is a lower bound on the $\ell_\infty$ norm, we have
\begin{align}
    \lambda_{max}(A^*A) &\le \|A^*A\|_\infty\\
    &\le \|A^*\|_\infty\|A\|_\infty\\
    &=\|A\|_1\|A\|_\infty.
\end{align}

Hence we prove $\|A\|_2 \le \sqrt{\|A\|_1\|A\|_\infty}$.

\hfill $\square$
\end{proof}

Now we sort $\{B_{\bot y}:y\in S\}$ as $p_1\ge p_2\ge\cdots\ge p_k$, in which $k=|S|=m+2$. Then we have
\begin{align}
\|W\|_1 &= \max_{y\in S} \sum_{x\in S}|W_{xy}| \\
&= \max_{1\le j\le k}\Big\{1+(k-1)\frac{1+p_j}{1-p_j}\Big\}\\
&= 1+ (k-1)\frac{1+p_1}{1-p_1}\\
&= k+(2k-2)\frac{p_1}{1-p_1},
\end{align}
and
\begin{align}
\|W\|_\infty &= \max_{x\in S} \sum_{y\in S}|W_{xy}| \\
&= \max_{1\le i\le k}\Big\{1+\sum_{j\ne i}\frac{1+p_j}{1-p_j}\Big\}\\
&\le \sum_{i=1}^k\frac{1+p_j}{1-p_j}\\
&=k + 2\sum_{i=1}^k\frac{p_j}{1-p_j}\\
&\le k + 2\sum_{i=1}^k\frac{p_j}{1-p_1}\\
&= k+\frac{2}{1-p_1}.
\end{align}

Therefore, we have 
\begin{equation}
    \|W\|_2 \le \sqrt{\Big(k+(2k-2)\frac{p_1}{1-p_1}\Big)\Big(k+\frac{2}{1-p_1}\Big)}
\end{equation}
and
\begin{equation}
    K_\delta = \|B^{-1}\|_2 \cdot \sqrt{\Big(k+(2k-2)\frac{p_1}{1-p_1}\Big)\Big(k+\frac{2}{1-p_1}\Big)}
\end{equation}
is feasible for $LP_2(B,B_\bot,0,1)$.

Similarly, denote $\tilde{W}$ as the matrix given by \eqref{eqn:lp1:solution}, then we have

\begin{align}
\|\tilde{W}\|_1 &= \max_{y\in S} \sum_{x\in S}|\tilde{W}_{xy}| \\
&= \max_{1\le j\le k}\Big\{1+(k-1)\frac{p_j}{1-p_j}\Big\}\\
&= 1+ (k-1)\frac{p_1}{1-p_1}
\end{align}
and
\begin{align}
\|\tilde{W}\|_\infty &= \max_{x\in S} \sum_{y\in S}|W_{xy}| \\
&= \max_{1\le i\le k}\Big\{1+\sum_{j\ne i}\frac{p_j}{1-p_j}\Big\}\\
&\le \max_{1\le i\le k}\Big\{1+\sum_{j}\frac{p_j}{1-p_j}\Big\}\\
&\le 1+\sum_{j}\frac{p_j}{1-p_1}\\
&= 1+ \frac{\sum_{j}p_j}{1-p_1}\\
&= 1+\frac{1}{1-p_1}
\end{align}

Therefore, we have 
\begin{equation}
    \|\tilde{W}\|_2 \le \sqrt{\Big(1+ (k-1)\frac{p_1}{1-p_1}\Big)\Big(1+\frac{1}{1-p_1}\Big)}
\end{equation}
and
\begin{equation}
    K_c = \|B^{-1}\|_2 \cdot \sqrt{\Big(1+ (k-1)\frac{p_1}{1-p_1}\Big)\Big(1+\frac{1}{1-p_1}\Big)}.
\end{equation}

Hence, $K_c+\delta K_\delta$ is feasible for $LP_1(B,B_\bot,c,\delta)$. Because our constructions for both parts make the equality hold in \eqref{eqn:lp1:cond3}, the final construction makes the equality hold naturally.

\subsection{Proof of Theorem~\ref{thm:scp}}
\label{app:scp}

For the $0$-IA property, according to Proposition~\ref{prop:colred} we only need to equivalently consider the case of $|\calM_*|+|\calC_*|$ malicious players in the canonical Byzantine setting. From Lemma~\ref{lem:byzantine}, the mechanism is $0$-IA even if up to $\frac{\delta}{2N}(n-1)$ malicious players are considered. Since $|\calM_*|+|\calC_*| \le \frac{\delta}{2N}(n-1)$, it is indeed $0$-IA.

Then we consider the colluding party. 
If all players in $\calC_*$ act honestly, since the mechanism is $(\delta,K)$-compact, each of them would get an interim utility of at least $\delta$ if there were no malicious players. As there are $\calM_*$ malicious players and each can perturb $r_{X_i}(Z_i)$ by at most $\frac{2N}{n-1}$, the actual interim utility of each player is at least $\delta-\frac{2N}{n-1}|\calM_*|$, so the total interim utility of the colluding party is
\begin{equation}
    u_{\calC_*}^{honest} \ge |\calC_*|\cdot\left(\delta-\frac{2N}{n-1}|\calM_*|\right).
\end{equation}

Assuming the mechanism is not weak-SCP, then there exists a case in which $1\le d \le |\calC_*|$ players in $\calC_*$ act dishonestly and increase the total interim utility of the colluding party. 
Hence, compared to the case that all players in $\calC_*$ act honestly, we can model this scenario as colluding players in $\calC_*$ change their actions, and consider the increment of their utilities.

As we assumed, $d$ players in $\calC_*'$ change their actions from honest to dishonest. Since there are now at most $|\calM_*|+d$ dishonest players, the interim utility of each player in $\calC_*'$ is at most $-\delta+\frac{2N}{n-1}(|\calM_*|+d)$; on the other hand, $d$ players changing their actions may increase the interim utility of each player in $\calC_*-\calC_*'$ by at most $\frac{2N}{n-1} d$. Hence, the increment of the total interim utility in $\calC_*$ is
\begin{align}
    \Delta &\le d\cdot \left((-\delta+\frac{2N}{n-1}(|\calM_*|+d))-(\delta-\frac{2N}{n-1}|\calM_*|)\right)+(|\calC_*|-d)\cdot \frac{2N}{n-1} d\\
    &=\left(-2\delta + \frac{2N}{n-1}(2|\calM_*|+|\calC_*|)\right)d\\
    &\le \left(-2\delta + \frac{4N}{n-1}(|\calM_*|+|\calC_*|)\right)d\\
    &= \left(-2\delta + \frac{4N}{n-1}\cdot \frac{\delta}{2N}(n-1)\right)d\\
    &=0.
\end{align}

Therefore, we show that the deviation cannot increase the colluding party's total utility, i.e. the mechanism is SCP when $|\calM_*|+|\calC_*| \le \frac{\delta}{2N}(n-1)$.

\subsection{Proof of Theorem~\ref{thm:cost}}
\label{app:thm:cost}

From Theorem~\ref{thm:Nbound} we see that for any $\delta\ge 0$, $LP_1$ has a feasible solution with the equality in Eq.~\eqref{eqn:lp1:cond3} holding and objective value
\begin{equation}
    K \le \|B^{-1}\|_2(c_1\cdot g_1(k,p_1)+\delta\cdot g_2(k,p_1)).
\end{equation}
To ensure a compactness of at least $\eta$, we only need $\delta \ge \eta K$.

In fact, assuming $\eta < \frac{1}{ g_2(k,p_1)\|B^{-1}\|_2}$, if we let $\delta = \frac{\eta c_1g_1(k,p_1)\|B^{-1}\|_2}{1-\eta g_2(k,p_1)\|B^{-1}\|_2}$ as in Eq.~\eqref{eqn:mu}, then
\begin{align}
    \delta - \eta K &\ge  \frac{\eta c_1g_1(k,p_1)\|B^{-1}\|_2}{1-\eta g_2(k,p_1)\|B^{-1}\|_2} - \eta \cdot \|B^{-1}\|_2(c_1g_1(k,p_1)+\delta\cdot g_2(k,p_1))\\
    &= \frac{\eta c_1g_1(k,p_1)\|B^{-1}\|_2-\eta \cdot \|B^{-1}\|_2(c_1g_1(k,p_1)+\delta g_2(k,p_1))(1-\eta g_2(k,p_1)\|B^{-1}\|_2) }{1-\eta g_2(k,p_1)\|B^{-1}\|_2}\\
    &= \frac{-\eta \|B^{-1}\|_2\cdot\left(-c_1g_1(k,p_1) \eta g_2(k,p_1)\|B^{-1}\|_2+\delta g_2(k,p_1)-\eta\delta (g_2(k,p_1))^2\|B^{-1}\|_2\right)}{1-\eta g_2(k,p_1)\|B^{-1}\|_2}\\
    &= \frac{-\eta g_2(k,p_1)\|B^{-1}\|_2\cdot(-\eta c_1g_1(k,p_1)\|B^{-1}\|_2+\delta\cdot (1-\eta g_2(k,p_1)\|B^{-1}\|_2))}{1-\eta g_2(k,p_1)\|B^{-1}\|_2}\\
    &= \frac{-\eta g_2(k,p_1)\|B^{-1}\|_2}{1-\eta g_2(k,p_1)\|B^{-1}\|_2}\cdot (-\eta c_1g_1(k,p_1)\|B^{-1}\|_2 + \eta c_1g_1(k,p_1)\|B^{-1}\|_2)\\
    &=0.
\end{align}
Hence, the $\eta$ compactness is satisfied. 

Now we only need to show that  $\mu = \delta$. Actually, whenever a verifier gets an observation $x$ she pays the cost of $c(x)$, and because the equality holds in Eq.~\eqref{eqn:lp1:cond3}, she gets an expected reward of $c(x)+\delta$ over the (conditional) distribution of other verifiers' observations. Hence we see that the expected payment to any verifier is $\delta$ plus the expected verification cost, and that $\mu=\delta$ holds indeed.

\subsection{Proof of Lemma~\ref{lem:tv}}
\label{app:lem:tv}

We only need to prove that 
$$\left|\sum_{X_j\in S}P(X_j|X_i,\hat{\phi}) T_{Z_iX_j} - \sum_{X_j\in S}P(X_j|X_i,\phi) T_{Z_iX_j}\right| \le \delta.$$

In fact,
\begin{align}
&\quad\left|\sum_{X_j\in S}P(X_j|X_i,\hat{\phi}) T_{Z_iX_j} - \sum_{X_j\in S}P(X_j|X_i,\phi) T_{Z_iX_j}\right| \nonumber\\
&= \left|\sum_{X_j\in S}T_{Z_iX_j} \left(P(X_j|X_i,\hat{\phi}) - P(X_j|X_i,\phi)\right) \right|\\
&\le \sum_{X_j\in S}|T_{Z_iX_j}|\cdot\left| P(X_j|X_i,\hat{\phi}) - P(X_j|X_i,\phi)\right|\\
&\le \sum_{X_j\in S}N\cdot\left| P(X_j|X_i,\hat{\phi}) - P(X_j|X_i,\phi)\right|\\
&= N\cdot\sum_{X_j\in S}\left| P(X_j|X_i,\hat{\phi}) - P(X_j|X_i,\phi)\right|.
\end{align}

From the definition of TV distance, we have 
\begin{align}
&TV_{X_j}(P(X_j|X_i,\hat{\phi}),P(X_j|X_i,\phi))\nonumber\\
&= \frac{1}{2}\sum_{X_j\in S}\left| P(X_j|X_i,\hat{\phi}) - P(X_j|X_i,\phi)\right|.
\end{align}

Hence,
\begin{align}
&N\cdot\sum_{X_j\in S}\left| P(X_j|X_i,\hat{\phi}) - P(X_j|X_i,\phi)\right| \nonumber\\
&= 2N\cdot TV_{X_j}(P(X_j|X_i,\hat{\phi}),P(X_j|X_i,\phi))\\
&\le 2N\cdot \frac{\delta}{2N}
\\&=\delta.
\end{align}

Here we prove Lemma~\ref{lem:tv}.

\subsection{Proof of Theorem~\ref{thm:inacc:prior}}
\label{app:proof:prior}

From Theorem~\ref{thm:generalization}, without loss of generality we consider the $2$-verifier DVG. Because environments $\phi,\hat{\phi}$ are identical except for priors, we define the environmental constraint $\Phi$ as all environments identical to $\phi$ except for different priors. Then for $\varphi\in\Phi$, $P(X_i|\theta,\varphi)$ is a constant irrelevant to $\varphi$ and we regard $\Theta_s=P(\theta=s,\varphi)$ as the variable. We omit the $\varphi$ for simplicity in the following parts of the proof. Then, we can see that
\begin{equation}\label{eqn:belief:app}
P(X_{-i}|X_i)=\frac{\sum_{\theta}P(\theta)P(X_i|\theta)P(X_{-i}|\theta)}{\sum_{\theta}P(\theta)P(X_i|\theta)}
\end{equation}
is a function of $\Theta=\{\Theta_s\}$. We denote $\Q:S\to \R^S$ as:
\[
    \Q_{X_{-i}}(X_i,\Theta)=P(X_{-i}|X_i).
\]
Then, we derive stability of $\Q_{X_{-i}}(X_i,\cdot)$ via $\ell_1$-Lipschitz properties. While in Eq.~\eqref{eqn:belief:app} there is a natural constraint that $\sum_{s\in S} \Theta_s=1$, here we relax this constraint and allow arbitrary $\Theta\in \R^S$ for convenience in analysis. We have
\begin{align}
    \frac{\partial \Q_{X_{-i}}(X_i,\Theta)}{\partial \Theta_s}&=\frac{P(X_i|\theta=s)P(X_{-i}|\theta=s)}{\sum_{\theta}P(\theta)P(X_i|\theta)} - \frac{\sum_{\theta}P(\theta)P(X_i|\theta)P(X_{-i}|\theta)\cdot P(X_i|\theta=s)}{(\sum_{\theta}P(\theta)P(X_i|\theta))^2}\\
    &=\frac{P(X_i|\theta=s)P(X_{-i}|\theta=s)}{P(X_i)}-\frac{P(X_i,X_{-i})\cdot P(X_{i}|\theta=s)}{P^2(X_i)}.
\end{align}

Hence, for fixed $X_i$, it holds that
\begin{align}
    \sum_{X_{-i}\in S}\left|\frac{\partial \Q_{X_{-i}}(X_i,\Theta)}{\partial \Theta_s}\right|
    &\le\sum_{X_{-i}\in S}\left|\frac{P(X_i|\theta=s)P(X_{-i}|\theta=s)}{P(X_i)}\right| + \sum_{X_{-i}\in S}\left|\frac{P(X_i,X_{-i})\cdot P(X_{i}|\theta=s)}{P^2(X_i)}\right|\\
    &= \frac{P(X_i|\theta=s)}{P(X_i)}+\frac{P(X_i)\cdot P(X_{i}|\theta=s)}{P^2(X_i)}\\
    &=  \frac{2P(X_i|\theta=s)}{P(X_i)}\\
    &\le \frac{2}{P(X_i)}.
\end{align}

Therefore, we deduce that $\Q(X_i,\Theta)$ is $\frac{2}{P(X_i)}$-$\ell_1$-Lipschitz at point $\Theta$, in which $\frac{2}{P(X_i)}$ is a function of $\Theta$ because different priors result in different marginal probabilities of observations.

Now we denote that the priors in $\phi,\hat{\phi}$ as $\Theta_1,\Theta_2$, and consider the path $\omega:[0,1]\to \R^{S}$ from $\Theta_1$ to $\Theta_2$ as 
\[
\omega(t)=(1-t)\Theta_1+t\Theta_2,
\]
and denote the environmental variable corresponding to $\omega(t)$ as $\varphi_t$. Then, because $P(X_i)$ is a linear function of $\{P(\theta)\}$, it holds that: 
\begin{align}
P(X_i|\varphi_t)&=(1-t)P(X_i|\phi)+tP(X_i|\hat{\phi})\\
&\ge \min\{P(X_i|\phi),P(X_i|\hat{\phi})\}.
\end{align}

Hence, we deduce that $\Q(X_i,\cdot)$ is $\max\{\frac{2}{P(X_i|\phi)},\frac{2}{P(X_i|\hat{\phi})}\}$-$\ell_1$-Lipschitz on $\omega$.

For $X_i\in S^*$, from the assumption that $TV_\theta(P(\theta|\phi), P(\theta|\hat{\phi})) \le \frac{\delta}{4N}\cdot \min_{X_i\in S^*, \varphi \in \{\phi,\hat{\phi}\}} \left\{P(X_i|\varphi)\right\}$, we see that the $\ell_1$ length of $\omega$ is at most $\frac{\delta}{2N}\cdot \min_{X_i\in S^*, \varphi \in \{\phi,\hat{\phi}\}} \left\{P(X_i|\varphi)\right\}$. From the Lipschitz properties, we have 
\begin{equation}
    \|\Q(X_i,\Theta_1)-\Q(X_i,\Theta_2)\|_{1} \le \frac{\delta}{N}.
\end{equation}

Notice that the TV distance between two distributions is $\frac{1}{2}$ times the $\ell_1$ distance between the corresponding probability vectors, hence
\begin{equation}\label{eqn:tv2}
    TV_{X_{-i}}(P(X_{-i}|X_i,\phi),P(X_{-i}|X_i,\hat{\phi})) \le \frac{\delta}{2N}.
\end{equation}

For $X_i=1$, i.e. the observation is ``Dishonest'', from the assumption that $P(X_i=1|\theta\ne 1)=0$ we deduce that $X_i=1$ implies $\theta=1$. Hence, $P(X_{-i}|P(X_i=1))=P(X_{-i}|\theta=1)$ is a known constant (by the modeling in Section~\ref{sec:model}) and is not affected by the inaccurate prior distributions of $\theta$, so Eq.~\eqref{eqn:tv2} also holds for $X_i=1$.

According to Theorem~\ref{thm:tv}, Eq.~\eqref{eqn:tv2} implies that the mechanism is $0$-IA for environment $\phi$.

\subsection{Proof of Theorem~\ref{thm:app:sscp}}
\label{app:sscp}

\noindent\textbf{(1)} From Eq.~\eqref{eqn:scp:scoring} we have
\begin{align}
    \sum_{i\in \calC}R_i(Z_i,\Z_{-i}) &= \sum_{i\in \calC} \frac{1}{n-1}\left( \sum_{j\ne i} Z_i'TZ_j\right)\\
    &=\frac{1}{n-1}\sum_{i\in \calC} \left(\sum_{j \notin \calC} Z_i'TZ_j +\sum_{j \in \calC \backslash \{i\}} Z_i'TZ_j\right) \\
    &=\frac{c(n-c)}{n-1}\overline{\Z_\calC}'T\overline{\Z_{-\calC}} + \frac{1}{n-1}\sum_{i\in \calC}\sum_{j \in \calC \backslash \{i\}} Z_i'TZ_j,
\end{align}
 in which $\overline{\Z_\calC}, \overline{\Z_{-\calC}}$ are the average reports of players inside and outside the colluding party $\calC$, respectively. Hence, we have
\begin{equation}
    \E_{\Z_{-\calC}\sim P(\X_{-\calC}|\X_{\calC})}\left[\sum_{i\in \calC}R_i(Z_i,\Z_{-i})\right] = \overline{\Z_\calC}'\cdot \frac{c(n-c)}{n-1} \E_{\Z_{-\calC}\sim P(\X_{-\calC}|\X_{\calC})}\left[T\overline{\Z_{-\calC}}\right]+ \frac{1}{n-1}\sum_{j \in \calC \backslash \{i\}} Z_i'TZ_j.
\end{equation}

We notice that for fixed $n, c$, scoring matrix $T$, and shared belief $P(\X_{-\calC}|\X_{\calC})$, the term 
$$\frac{c(n-c)}{n-1} \E_{\Z_{-\calC}\sim P(\X_{-\calC}|\X_{\calC})}\left[T\overline{\Z_{-\calC}}\right]$$
is a constant and $\overline{\Z_\calC}'\cdot \frac{c(n-c)}{n-1} \E_{\Z_{-\calC}\sim P(\X_{-\calC}|\X_{\calC})}\left[T\overline{\Z_{-\calC}}\right]$ is linear to $\overline{\Z_\calC}$, the \emph{average} of all players' reports in $\calC$. Hence, we have
\begin{align}
&\overline{\Z_\calC}'\cdot \frac{c(n-c)}{n-1} \E_{\Z_{-\calC}\sim P(\X_{-\calC}|\X_{\calC})}\left[T\overline{\Z_{-\calC}}\right] \le \max_{i\in \calC} \left\{\Z_i'\cdot \frac{c(n-c)}{n-1} \E_{\Z_{-\calC}\sim P(\X_{-\calC}|\X_{\calC})}\left[T\overline{\Z_{-\calC}}\right]\right\}.
\end{align}

Let $i^*$ be the $i\in\calC$ that yields the maximum in RHS. On the other hand, we have
\begin{equation}
\frac{1}{n-1}\sum_{i\in \calC}\sum_{j \in \calC \backslash \{i\}} Z_i'TZ_j = \frac{1}{n-1}\sum_{i\in \calC}\sum_{j \in \calC \backslash \{i\}} T_{Z_iZ_j}.
\end{equation}

Since $\forall T_{Z_iZ_j} \in [-N, N]$, 
\begin{equation}
\frac{1}{n-1}\sum_{i\in \calC}\sum_{j \in \calC \backslash \{i\}} Z_i'TZ_j \in \left[-\frac{c(c-1)}{n-1}N,\frac{c(c-1)}{n-1}N\right].
\end{equation}

Let $\forall Z_i=X_{i^*}$ and $h=\frac{2c(c-1)}{n-1}N$, we can see that Eq.~\eqref{eqn:scp:impossibility} holds. 

\vspace{1em}

\noindent\textbf{(2)}
We let $\forall Z_i = X_i$. Since for any $X_i\ne X_j$ it holds that $T_{X_iX_i}\ge T_{X_iX_j}$, and we first adopt the assumption that  $\exists i,j\in \calC$ s.t. $T_{X_iX_i}> T_{X_iX_j}$, we have
\begin{align}
\frac{1}{n-1}\sum_{i\in \calC}\sum_{j \in \calC \backslash \{i\}} Z_i'TZ_j &= \frac{1}{n-1}\sum_{i\in \calC}\sum_{j \in \calC \backslash \{i\}} T_{X_iX_j}\\
&< \frac{1}{n-1}\sum_{i\in \calC}\sum_{j \in \calC \backslash \{i\}} T_{X_iX_i}\label{eqn:scp:strict} \\  
&= \frac{c-1}{n-1}\sum_{i\in \calC} T_{X_iX_i}\\
&= \frac{c-1}{n-1}\sum_{i\in \calC} X_i' \cdot diag(T)\\
&=  \overline{\X_\calC}'\cdot \frac{c(c-1)}{n-1} diag(T).
\end{align}

Conditioned on the players in $\calC$ all actively verify and observe $X_\calC$, their observation costs are fixed, and their expected total reward is:
\begin{align}
r(\X_\calC) &= \overline{\X_\calC}'\cdot \frac{c(n-c)}{n-1} \E_{\X_{-\calC}\sim P(\X_{-\calC}|\X_{\calC})}\left[T\overline{\X_{-\calC}}\right]\nonumber\\
&~~~+ \frac{1}{n-1}\sum_{i\in \calC}\sum_{j \in \calC \backslash \{i\}} T_{X_iX_j}.
\end{align}

On the other hand, we let $k=|S|$ be the number of types and define $f: \R^{k} \to \R$ s.t.
\begin{align}
f(Y) &= Y'\cdot \frac{c(n-c)}{n-1} \E_{\X_{-\calC}\sim P(\X_{-\calC}|\X_{\calC})}\left[T\overline{\X_{-\calC}}\right]\nonumber\\
&~~~+ Y'\cdot \frac{c(c-1)}{n-1} diag(T).
\end{align}
From the arguments above, it holds that
\begin{equation}
r(\X_\calC) < f(\overline{\X_\calC}).
\end{equation}

Since $f(\cdot)$ is a linear function, we have 
\begin{equation}
    f(\overline{\X_\calC}) \le \max_{i\in \calC} f(X_i).
\end{equation}

Let $i^*= \arg\max_{i\in \calC}f(X_i)$, then we deduce that
\begin{equation}
    r(\X_\calC) < f(X_{i^*}).
\end{equation}

Therefore, for the all-same report $\Z_{\calC}^* = \{X_{i^*},\cdots,X_{i^*}\}$, we have

\begin{align}
r(\Z_{\calC}^*) &= \overline{\Z_{\calC}^*}'\cdot \frac{c(n-c)}{n-1} \E_{\X_{-\calC}\sim P(\X_{-\calC}|\X_{\calC})}\left[T\overline{\X_{-\calC}}\right] + \frac{1}{n-1}\sum_{i\in \calC}\sum_{j \in \calC \backslash \{i\}} T_{Z_i^*Z_j^*}\\
&= X_{i^*}'\cdot \frac{c(n-c)}{n-1} \E_{\X_{-\calC}\sim P(\X_{-\calC}|\X_{\calC})}\left[T\overline{\X_{-\calC}}\right] + \frac{1}{n-1}\sum_{i\in \calC}\sum_{j \in \calC \backslash \{i\}} T_{X_{i^*}X_{i^*}}\\
&=X_{i^*}'\cdot \frac{c(n-c)}{n-1} \E_{\X_{-\calC}\sim P(\X_{-\calC}|\X_{\calC})}\left[T\overline{\X_{-\calC}}\right] + X_{i^*}'\cdot \frac{c(c-1)}{n-1} diag(T)\\
&=f(X_{i^*})\\
&>r(\X_\calC).
\end{align}

Hence we show that such collusion is strictly profitable in the shared-belief setting, implying that the mechanism is not strong-SCP.

If we do not adopt the assumption that $\exists i,j\in \calC$ s.t. $T_{X_iX_i}> T_{X_iX_j}$, then Eq.~\eqref{eqn:scp:strict} still holds with ``$\le$'', and all the following strict inequalities become non-strict, implying that Eq.~\eqref{eqn:scp:impossibility} holds with $h\le 0$.
\end{document}